\documentclass{article}
\PassOptionsToPackage{numbers,compress,sort&compress}{natbib}

\usepackage[preprint]{neurips_2026}

\usepackage[utf8]{inputenc}
\usepackage[T1]{fontenc}
\usepackage{hyperref}
\usepackage{url}
\usepackage{booktabs}
\usepackage{amsfonts}
\usepackage{amsmath}
\usepackage{nicefrac}
\usepackage{microtype}
\usepackage{graphicx}
\usepackage{multirow}
\usepackage{pifont}
\usepackage{kotex}  
\usepackage{adjustbox}
\usepackage[table,xcdraw]{xcolor}
\usepackage{graphicx}              
\usepackage{subcaption}
\usepackage{longtable}
\usepackage{wrapfig}
\usepackage{algorithm}
\usepackage{algpseudocode}
\usepackage{amssymb}
\usepackage{wrapfig}
\usepackage{soul}
\usepackage[most]{tcolorbox}
\usepackage{inconsolata}   

\usepackage{caption}

\definecolor{hlyellow}{HTML}{FFF59D}   
\definecolor{hlgreen}{HTML}{D9F2D7}    
\definecolor{hlpink}{HTML}{FFD5E5}     
\definecolor{hlblue}{HTML}{CFE8FF}     

\newcommand{\hlp}[1]{\sethlcolor{hlpink}\hl{#1}}
\newcommand{\hlb}[1]{\sethlcolor{hlblue}\hl{#1}}

\sethlcolor{tokenbg}

\AtBeginEnvironment{longtable}{\ttfamily\footnotesize}
\usepackage{dirtree}

\title{Remember Your Trace: Memory-Guided Long-Horizon Agentic Framework for Consistent and Hierarchical Repository-Level Code Documentation}

\author{
 \textbf{Suyoung Bae\textsuperscript{1}},
 \textbf{Jaehoon Lee\textsuperscript{1}},
 \textbf{Changkyu Choi\textsuperscript{2}}
 \textbf{YunSeok Choi\textsuperscript{1}\thanks{Corresponding authors}},
 \textbf{Jee-Hyong Lee\textsuperscript{1}$^*$} \\
 \textsuperscript{1} Sungkyunkwan University, South Korea \\
 \textsuperscript{2} University of Oslo, Norway \\
}

\begin{document}

\maketitle

\begin{abstract}
Automated code documentation is essential for modern software development, providing the contextual grounding that both human developers and coding agents rely on to navigate large codebases. Existing repository-level approaches process components independently, causing redundant retrieval and conflicting descriptions across documents while producing outputs that lack hierarchical structure. Therefore, we propose \textbf{MemDocAgent}, a long-horizon agentic framework that generates documentation within a single, integrated context spanning the entire repository. It combines two components: (i) \textit{Dependency-Aware Traversal Guiding} that predetermines a traversal order respecting dependency and granularity hierarchies; (ii) \textit{Memory-Guided Agentic Interaction}, in which the agent interacts with RepoMemory, a shared memory accumulating prior work traces through read, write, and verify operations. Through an in-depth multi-criteria evaluation, MemDocAgent achieves the best performance over both open- and closed-source baselines and demonstrates practical applicability in real software development workflows.~\footnote{Code and Data are available at \url{https://github.com/bsy99615/MemDocAgent}.}
\end{abstract}

\section{Introduction}

Repository-level documentation requires understanding over an entire codebase to generate coherent natural-language descriptions that span individual function behaviors, dependency structures, and overall architectural patterns~\citep{docagent, codewiki, repoagent}.
As software repositories scale in size and complexity, such documentation serves as a critical interface for both human developers and coding agents, providing the contextual grounding needed for code navigation and collaboration~\citep{Parnas2010PreciseDT, GAROUSI2015664, kivinen2023ai}. 
Manual writing at this scale is prohibitively expensive, motivating a {new} line of work on automating repository-level documentation~\citep{parnas1997software, forward2002relevance, de2005study, Aghajani2019SoftwareDI, zhi2015cost, xia2017measuring}. 

\begin{figure}[t]
\centering
\includegraphics[width=1\linewidth]{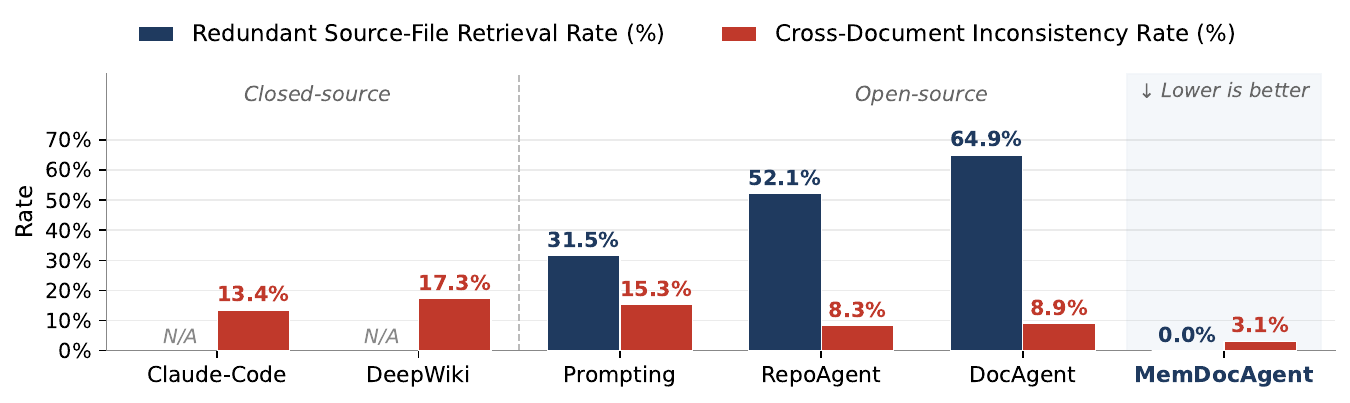}
\caption{Limitations in existing documentation systems. While existing systems suffer from substantial repeated source-file retrieval and cross-document inconsistencies, MemDocAgent eliminates repeated retrieval and reduces inconsistency by 75.5\%. \textit{N/A} denotes closed-source systems whose internal pipelines are inaccessible, so repeated retrieval cannot be measured. The cross-document inconsistency rate is computed post hoc over the final document set by a separate LLM-judge (Claude Haiku 4.5), applied identically to all systems: it is the fraction of dependency-linked document pairs whose descriptions the judge deems contradictory.}
\label{fig:figure1}
\end{figure}

Automating this task is inherently a long-horizon: each document depends not only on the local code being summarized, but also on information retrieved, interpreted, and written in earlier documentation steps. 
%
%
However, existing repository-level documentation systems, ranging from open-source frameworks~\citep{repoagent, docagent, codewiki} to closed-source products~\citep{deepwiki, claudecode}, decompose a repository into components and document each component independently, \emph{treating each as a self-contained task without shared state across the repository}. As a result, the same source files are repeatedly retrieved across iterations, {with an average overlap of 50\% across existing systems} as shown in Figure~\ref{fig:figure1}. More critically, prior documentation outputs are not reused across iterations, leaving the system unable to detect dependency conflicts.
{This leads to contradictory descriptions of}
the same entity or architectural logic across documents, resulting in an average cross-document inconsistency rate of 13\% across existing systems as shown in Figure~\ref{fig:figure1}. 

Beyond these procedural limitations, existing methods \emph{lack hierarchical structure in their documentation outputs}. High-quality repository documentation should provide a multi-level view of a codebase, connecting fine-grained component behavior, module-level responsibilities, and repository-wide architectural design into a coherent whole~\citep{garousi2013evaluating}. However, existing methods typically document only a narrow slice of this hierarchy: component-level approaches generate localized docstrings but overlook architectural context~\citep{repoagent, docagent}, whereas repository-level approaches summarize global structure but omit fine-grained implementation details~\citep{codewiki}. Consequently, the resulting documentation is either too local to explain system-level design or too coarse to support concrete code understanding. This hierarchical gap reinforces the need for a long-horizon documentation process that can accumulate, preserve, and connect information across many interdependent documentation steps.


Existing systems treat repository documentation as short-horizon, component-local tasks, whereas the task requires accumulating and reconciling repository-wide knowledge across many documentation steps. 
{We therefore formulate repository-level documentation as a \emph{long-horizon agentic problem}~\citep{erdogan2025plan}, where a single agent maintains persistent state, reuses prior retrievals and outputs, and builds a coherent hierarchical documentation of the codebase.}
%
%
To this end, we propose \textbf{MemDocAgent}, a long-horizon agentic framework in which a single agent reasons over the entire repository within one continuous trajectory, with persistent access to its accumulated retrievals and outputs. Along this trajectory, the agent produces hierarchical documentation spanning the component, module, and repository levels. This design is supported by two core components: \textit{Dependency-Aware Traversal Guiding} 
and \textit{Memory-Guided Agentic Interaction}
. 


\textit{Dependency-Aware Traversal Guiding} predetermines a traversal order that respects the dependency relations and the granularity hierarchy, and provides it as guidance to the agent to ensure hierarchical structure and coverage completeness. In \textit{Memory-Guided Agentic Interaction}, the agent interacts with RepoMemory, a shared memory that accumulates prior work traces such as retrieved components, intermediate reasoning, and generated documentation throughout the trajectory. 
Crucially, RepoMemory is not only a database that passively accumulates generated documentation, but also an active manager that governs the documentation process. Beyond maintaining hierarchical completeness, it serves as a consistency-enforcement mechanism, actively keeping interdependent documents consistent with one another.

To validate the effectiveness of MemDocAgent, we comprehensively evaluate it against existing baselines along four dimensions: \textit{completeness}, \textit{truthfulness}, \textit{helpfulness}, and \textit{information sufficiency}. Existing metrics capture coverage, accuracy, and usefulness, but overlook implementation-level sufficiency for practical code understanding. We therefore introduce information sufficiency as a quantitative metric that measures whether the generated documentation alone provides enough information to reproduce the original source code.
%
Empirically, MemDocAgent achieves the best performance across all four criteria, demonstrating that its documentation captures both the breadth and the depth required to support real software development workflows.

\section{Related work}

\paragraph{Long-horizon coding agents.}
Recent coding agents have advanced from solving isolated programming tasks to executing extended trajectories with numerous sequential tool calls and reasoning steps~\citep{yang2024swe, wang2024openhands, zhang2024codeagent, phan2024hyperagent, bouzenia2025repairagent, deng2025swe, luo2025ultrahorizon, ye2025agentfold, liu2025context, gao2026swe, kim20262agent}. This long-horizon regime builds on agentic foundations such as ReAct~\citep{yao2022react}, Reflexion~\citep{shinn2023reflexion}, and multi-agent frameworks~\citep{hong2023metagpt, qian2024chatdev, wu2024autogen}. A key challenge is that working memory accumulates across turns and exceeds the context window, causing agents to lose track of prior objectives or propagate errors throughout the trajectory~\citep{liu-etal-2024-lost, sinhaillusion, wan2025compass, ning2026code}.

To handle long contexts, prior work has explored external memory systems that maintain and dynamically organize context outside the model~\citep{park2023generative, packer2023memgpt, wang2023voyager, zhong2024memorybank, gutierrez2024hipporag, xu2025amem, hu2025hiagent, chen2025swe, mu2026experepair, wang2026improving}, along with summarization and compression based techniques for context management~\citep{wu2025resum, kang2025acon, li2025sculptor}. 
However, these methods are designed for single output settings where memory supports a single final answer, and fall short when multiple sequentially generated outputs must remain mutually consistent. 
We instead reformulate repository documentation as generating mutually consistent documents over one cumulative trajectory, where memory must preserve cross-document consistency rather than support a single answer.

\paragraph{Repository-level code documentation.}
Early code documentation methods focused on function-level summarization, training models to produce short natural language descriptions for individual code units~\citep{moreno2013automatic, iyer-etal-2016-summarizing, leclair2019recommendations, choi2023readsum, poudel2024documint}. With the emergence of LLMs, this scope expanded to summarization at the class and project levels~\citep{khan2022automatic, lomshakov-etal-2024-proconsul, makharev2025code, geng2024large, ahmed2024automatic, song2024code, cao-etal-2025-rethinking}, and recent systems have begun to operate over entire repositories. RepoAgent~\citep{repoagent} extracts dependency relations and generates component-level docstrings in topological order through an LLM pipeline, while DocAgent~\citep{docagent} adopts a multi-agent pipeline for the same component-level objective. CodeWiki~\citep{codewiki}, in contrast, hierarchically decomposes a repository and recursively synthesizes documentation via a multi-agent pipeline targeting repository-level abstractions. 

While these methods successfully extend documentation generation to the repository scale, they treat each component as an independent unit, thereby failing to maintain cross-document consistency or dependency-aware reasoning across the long-horizon documentation trajectory.
This per-component processing leads to redundant retrievals and inter-document contradictions that undermine the practical utility of the resulting documentation.

\section{Preliminaries}
\paragraph{Repository-level hierarchical documentation.}
Given a repository $\mathcal{R}$ with a hierarchical directory structure and multiple source files, our task is to produce a hierarchical documentation set $\mathcal{D}$ over three granularities: \emph{components} (functions, methods, classes), \emph{modules} (directories), and the \emph{repository} itself. The corresponding documentation units are:
\begin{equation}
    \mathcal{U} \;=\; \{u^{(c)}_i\}_{i=1}^{N_c} \;\cup\; \{u^{(m)}_j\}_{j=1}^{N_m} \;\cup\; \{u^{(r)}\},
\end{equation}
Where $N_c$ and $N_m$ are the numbers of components and modules, and the superscripts $(c)$, $(m)$, $(r)$ indicate the granularity level, respectively. 
The system processes every $u \in \mathcal{U}$ to generate a corresponding document $d \in \mathcal{D}$, yielding $|\mathcal{D}| = N_c + N_m + 1$ documents in total.

\paragraph{Long-horizon reasoning trajectories.}

We model the documentation generation process as a long-horizon reasoning trajectory, formally defined as $\tau = (t_0, a_0, o_0, t_1, a_1, o_1, \ldots, t_T, a_T, o_T)$, where each turn $i$ comprises a thought $t_i$, an action $a_i$, and an observation $o_i$ following the ReAct pattern~\citep{yao2022react}, and $T$ denotes the total number of turns in the trajectory. 
In repository-level documentation generation, a single trajectory typically spans over a thousand turns as the agent processes hundreds of components, dozens of modules, and the repository itself.


\begin{figure}[t]
\centering
\includegraphics[width=1\linewidth]{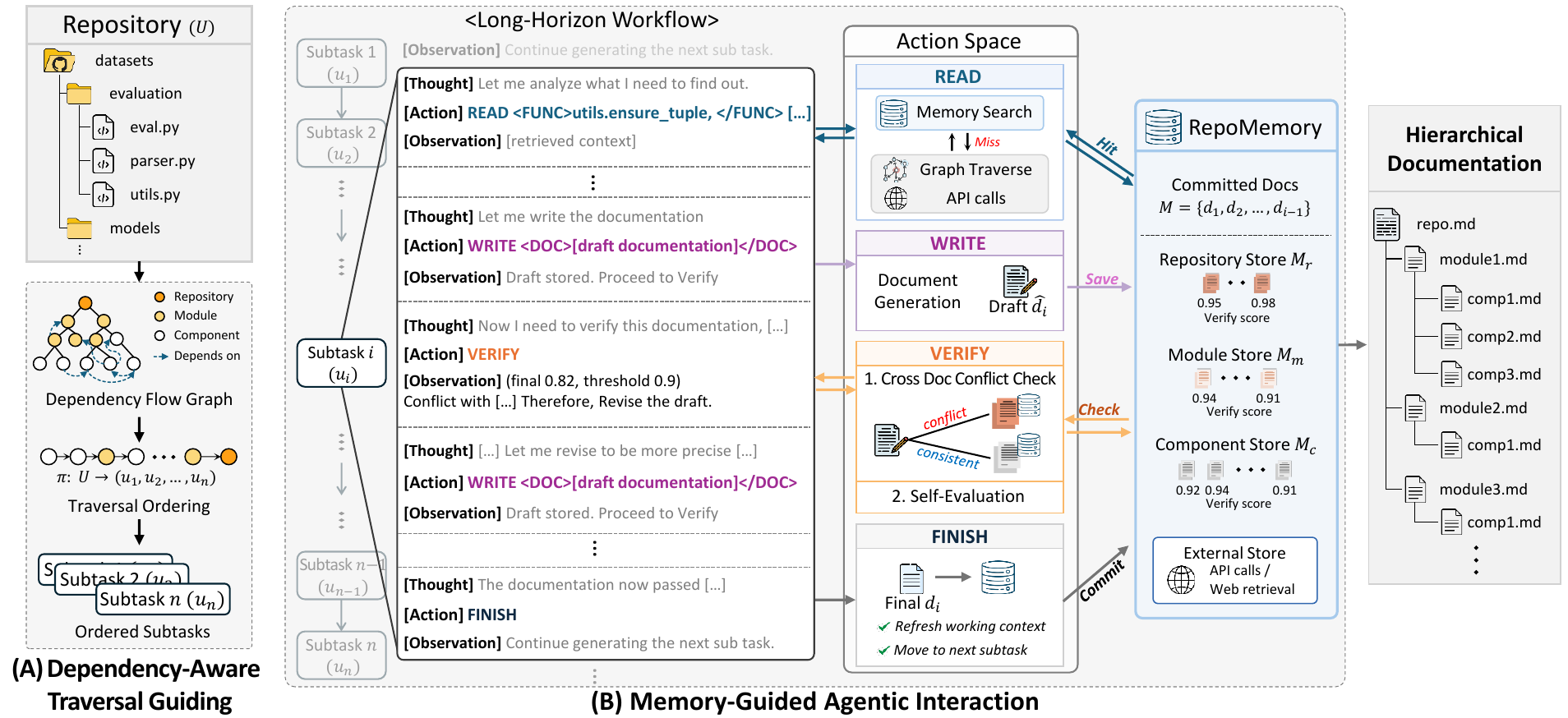}
\caption{Overview of MemDocAgent. (A) \textit{Dependency-aware traversal guiding} first computes a traversal order that respects both dependency relations and the granularity hierarchy. Following this order, (B) \textit{memory-guided agentic interaction} treats each unit as a multi-turn sub-task, where the agent interacts with RepoMemory through \textsc{Read}, \textsc{Write}, and \textsc{Verify} to generate $d_i$ and commits the verified document upon \textsc{Finish}. The local context is refreshed before moving to the next sub-task, and this loop repeats until all units are documented, producing the final hierarchical documentation.}
\vspace{-0.6em}
\label{fig:main_figure}
\end{figure}

\section{MemDocAgent}


MemDocAgent frames repository-level documentation as a cumulative long-horizon process, where a single agent processes every documentation unit within one continuous trajectory and reuses information accumulated along the way. This design is supported by two key components. First, \textit{dependency-aware traversal} (\S\ref{sec:traversal}) that orders the trajectory so that each unit is documented after its dependencies and child units, ensuring both context-grounded generation and repository coverage. Second, \textit{memory-guided agentic interaction} (\S\ref{sec:workflow}), where the agent interacts with a shared external memory to retrieve, reuse, and update information across sub-tasks. Figure~\ref{fig:main_figure} illustrates our framework.

\subsection{Dependency-aware traversal guiding}~\label{sec:traversal}
\vspace{-1.5em}

To structurally guarantee the completeness of $\mathcal{D}$ for the target repository, {we predetermine a traversal order $\pi = (u_1, \ldots, u_N)$ over $\mathcal{U}$ and injects it into the agent as sequential sub-tasks. 
We first construct a directed dependency graph $G = (V, E)$ over $\mathcal{R}$, where each node $v \in V$ corresponds to a documentation unit and a directed edge $(u, v) \in E$ indicates that $u$ depends on $v$. 

Specifically, we traverse the abstract syntax tree (AST)~\citep{8812062} of each file to collect code entities (functions, methods and classes) as nodes, and connect them with directed edges representing their relations (function calls, class inheritance, attribute access and imports). Each node stores its AST type and code context, and each edge stores the dependency relation as metadata.

Unlike DocAgent~\citep{docagent}, which uses a topological order reflecting only component-level dependencies, $\pi$ jointly satisfies two constraints: (1) component-level dependency order, where $u_i$ precedes $u_j$ whenever $u_j$ depends on $u_i$; and (2) granularity-level hierarchical order, where each module is processed only after all of its components and sub-modules are documented, and the repository unit is processed only after all modules are complete. By jointly satisfying both, $\pi$ ensures that the agent always processes a unit with all of its dependent and child units already documented. Since code dependencies may contain cycles, we collapse each strongly connected component into a single super-node using Tarjan's SCC algorithm~\citep{tarjan1972depth} before topological ordering. Detailed graph construction and the traversal order algorithm are provided in Appendix~\ref{A.1}.

\subsection{Memory-guided agentic interaction} ~\label{sec:workflow}
\vspace{-1.5em}

To ensure both retrieval efficiency and cross-document consistency throughout the long-horizon trajectory, this section introduces a centralized shared memory, RepoMemory, and designs an interaction mechanism in which every action of the agent interacts with this memory.


\begin{table}[t]
\caption{The structure of RepoMemory.}
\label{tab:repomemory}
\centering
\resizebox{\linewidth}{!}{%
\begin{tabular}{llll}
\toprule
Memory store &  & Key & Value \\
\midrule
\multirow{1}{*}{Documentation store} & component\_store & component\_id & path, document, claims, depends\_on, source code, type, verification\_score \\
& module\_store & module\_path & path, document, claims, child\_units, verification\_score \\
& repo\_store & repo\_name & path, document, claims, child\_units, verification\_score \\
\multirow{1}{*}{External store} & search\_cache & query & search result \\
\bottomrule
\end{tabular}%
}
\vspace{-0.4em}
\end{table}

\subsubsection{RepoMemory}

RepoMemory consists of a \textit{Documentation store} and an \textit{External store} as shown in Table~\ref{tab:repomemory}. The documentation store organizes generated documents by hierarchy level, storing component, module, and repository-level records with their identifiers, paths, documents, decomposed claims, verification scores, source code and type (function, method, class) for components, child units for module and repository units. The external store caches search results by query to avoid redundant retrievals during the trajectory.


\subsubsection{Action space}

Unlike prior repository-level documentation systems whose actions are executed independently, every action in MemDocAgent is grounded in interaction with RepoMemory: \textsc{Read} retrieves context from RepoMemory, \textsc{Write} produces a document using the retrieved context, \textsc{Verify} cross-validates the generation against documents already committed to RepoMemory, and \textsc{Finish} commits the verified document to RepoMemory. At each turn, the agent observes the result of its previous action and autonomously selects the next action among these four operations to complete the current documentation sub-task. All instructions for each action are defined in the agent's system prompt (Appendix~\ref{A.2} for the full system prompt and Appendix~\ref{A.5} for trajectory examples).

\paragraph{\textsc{Read}: Memory-guided adaptive retrieval.}
Through \textsc{Read}, the agent adaptively decides whether additional context is needed for each sub-task $u_i$. If so, it calls the retrieval tool with two types of structured requests: \textit{internal requests} that specify related component identifiers within the repository, and \textit{external requests} that pose natural-language queries about external algorithms or libraries.

The retrieval tool first checks RepoMemory by key: component IDs for internal requests and query strings for external requests. If the key exists, the stored value is returned at $O(1)$ cost; otherwise, the tool retrieves the information from the codebase via the dependency graph $\mathcal{G}$ or from the external retrieval API, and stores it for later reuse.
For module and repository-level sub-tasks, the retrieval tool additionally retrieves the documentation of all direct child units from the \textit{Documentation Store}. Since the traversal order $\pi$ guarantees that these children have already been documented, this step reuses existing documentation without additional codebase retrieval.



\paragraph{\textsc{Write}: Hierarchical document generation.}
When the agent thinks that sufficient context has been collected, it performs \textsc{Write} to generate a draft document $\hat{d}_i$. The generation is guided by a granularity-specific format, defined in the agent's system prompt, that specifies the structure and required content of the output for each granularity.

\paragraph{\textsc{Verify}: Self-evaluation and cross-document conflict verification.}
Repository-level documentation often describes the same functionality from multiple perspectives across files, so without verification these descriptions easily drift apart and degrade overall consistency. Therefore, \textsc{Verify} assesses each draft $\hat{d}_i$ along two axes. First, a self-evaluation scores factual consistency, completeness, and helpfulness according to the agent's system prompt instructions (See Table~\ref{tab:example_1} in Appendix~\ref{A.2}), and second, a cross-conflict verification scores inconsistencies between $\hat{d}_i$ and its dependency-related documents, each on $[0, 1]$.    

For the cross-conflict verification, inspired by NLI-based inconsistency detection for text summarization~\citep{laban2022summac, scire2024fenice, yang-etal-2024-fizz}, we decompose $\hat{d}_i$ into atomic claims via an LLM and verify each claim with a local NLI model against a trusted reference set from RepoMemory. This reference set contains only the dependency-related records whose verification score exceeds a threshold, so a low-quality document never becomes a conflict-checking criterion and early errors cannot propagate. The resulting contradiction rate serves as the cross-conflict score.
$\hat{d}_i$ passes \textsc{Verify} only if the average of two verified scores exceeds a threshold. Otherwise, the detected conflicts are returned as a conflict report in the next turn's observation, prompting the agent to issue an additional \textsc{Read} or to re-\textsc{Write} the draft. If $\hat{d}_i$ still fails to pass after the agent has reached its predefined maximum number of revisions, we commit the highest-scoring draft to RepoMemory to preserve coverage. More details are provided in Appendix~\ref{A.4}.




\paragraph{\textsc{Finish}: Memory commit and trajectory continuation.}
Once $\hat{d}_i$ passes \textsc{Verify}, the agent finalizes it as $d_i$, invokes \textsc{Finish} to commit $d_i$ to the corresponding granularity store, and refreshes its local context for the next sub-task. This loop repeats until all units in $\mathcal{U}$ are documented, yielding the final hierarchical documentation set $\mathcal{D}$.

\section{Experimental setups}

\subsection{Evaluation metrics}

    


We evaluate MemDocAgent along four criteria. Beyond \textit{completeness}, \textit{truthfulness}, and \textit{helpfulness}, which are established quality indicators, we further propose \textit{information sufficiency}. This criterion quantifies whether the documentation contains enough information to reproduce source code that integrates and executes coherently within the repository. We outline the evaluation metrics below; full details are in Appendix~\ref{B.1}.

\paragraph{Completeness.}
Completeness measures the extent to which the generated documentation adheres to standard structural conventions (\textit{Section Presence}) and includes essential components (e.g., function, class) expected for a given code element (\textit{Entity Coverage}). The final completeness score is computed as the average of the section presence and entity coverage scores.

\textit{Section Presence Score.} For each documentation, we measure the proportion of required sections that are present, using rule-based pattern matching. This score is a sanity check that verifies whether MemDocAgent actually produces the intended hierarchical structure with all prescribed sections.


\textit{Entity Coverage Score.} We extract the set of core entities from the target source code via AST analysis, and measure the proportion of these entities that are referenced in the documentation. 
The core entities depend on the granularity level; At the component level, they are the function, class, method, and variable names. At the module level, they are the names of child components and sub-modules. At the repository level, they are the names of top-level modules and of the components along their immediate paths.

\paragraph{Truthfulness.}
Truthfulness measures whether the generated documentation faithfully reflects the actual code, covering both factual correctness (\textit{Consistency}) and contextual relevance (\textit{Relevance}).
We adopt a fine-grained documentation evaluation framework that decomposes generated documentation and evaluates each segment using predefined evaluation criteria~\citep{min-etal-2023-factscore, bae-etal-2026-referee} to compute these two scores. The final truthfulness score is computed as the product of the consistency and relevance scores. 

\textit{Consistency Score.} We measure the proportion of claims in the documentation that are factually supported by the source code, identified at fine-grained segment level.

\textit{Relevance Score.} We measure the proportion of documentation content that pertains to the source code, filtering out off-topic or extraneous descriptions.

\paragraph{Helpfulness.}
We assess the semantic quality and practical utility of each documentation through an LLM-as-a-judge protocol following DocAgent~\citep{docagent}. The judge rates each required section on a 5-point Likert scale using a section-specific rubric, and the normalized section scores are averaged into a single helpfulness score.

\paragraph{Information Sufficiency.}
Surface-level documentation quality alone does not capture whether the documentation is actually usable in practice. Motivated by the real-world scenario where documentation serves as a context tool for coding agents, we introduce an execution-based metric that measures whether the documentation contains sufficient information to support practical code reconstruction.
We operationalize this by measuring code regeneration performance: if a model can regenerate the original component using only the documentation and the function signature, the documentation is likely to preserve the behavior, dependencies, and implementation details needed for downstream use. 

%
%
To evaluate information sufficiency, we collect 564 component-level test cases from DevEval~\citep{deveval}. The evaluation proceeds as follows: (1) remove each target function body while keeping its signature; (2) provide the signature and documentation context to a code-generation model; (3) regenerate the missing function body; and (4) reinsert the generated code into the repository and evaluate it with functional correctness (\textit{Pass@k}) and structural similarity (\textit{CodeBLEU}).

%





\begin{table*}[t]
\caption{Main comparison results. (a) Overall results on average completeness, truthfulness, and helpfulness and (b) the same metrics separated by documentation granularity. Within each backbone group, the best result is \textbf{bold} and the second-best \underline{underlined}. * marks in section scores set to 1.0 by default, as their internal format instruction is inaccessible. ``--'' denotes non-applicable granularities.}
\label{tab:main_result}
\centering

\begin{subtable}[t]{\linewidth}
\centering
\caption{Overall results on average completeness, truthfulness, and helpfulness.}
\label{tab:main_result_overall}
\resizebox{\linewidth}{!}{%
\begin{tabular}{l l c ccc ccc c}
\toprule
\textbf{Backbone} & \textbf{Method} & \#Doc
& \multicolumn{3}{c}{\textbf{Completeness}} & \multicolumn{3}{c}{\textbf{Truthfulness}} & \\
\cmidrule(lr){4-6} \cmidrule(lr){7-9}
& & 
& \begin{tabular}[c]{@{}c@{}}Section\\Score\end{tabular}
& \begin{tabular}[c]{@{}c@{}}Coverage\\Score\end{tabular}
& Complete.
& Consistency & Relevance & Truthful.
& \textbf{Helpfulness} \\
\midrule
\multicolumn{9}{c}{\textit{Open-Source Baselines}} \\
\midrule
\multirow{5}{*}{Qwen3-Coder}
        & Prompting    & 3,323 & \underline{0.838} & 0.837 & 0.838 & 0.809 & 0.803 & 0.649 & 0.458 \\
  &     RepoAgent    & 3,650 & 0.659 & \underline{0.967} & 0.813 & \underline{0.910} & \underline{0.918} & \underline{0.835} & 0.391 \\
  &     DocAgent     & 3,198 & 0.792 & 0.898 & \underline{0.845} & 0.900 & 0.889 & 0.800 & 0.556 \\
  &     CodeWiki     & 80    & 0.770 & 0.473 & 0.622 & 0.906 & 0.852 & 0.772 & \underline{0.628} \\
  &     \cellcolor[HTML]{EFEFEF}\textbf{MemDocAgent}
        & \cellcolor[HTML]{EFEFEF}3,323
        & \cellcolor[HTML]{EFEFEF}\textbf{0.959}
        & \cellcolor[HTML]{EFEFEF}\textbf{0.998}
        & \cellcolor[HTML]{EFEFEF}\textbf{0.979}
        & \cellcolor[HTML]{EFEFEF}\textbf{0.968}
        & \cellcolor[HTML]{EFEFEF}\textbf{0.946}
        & \cellcolor[HTML]{EFEFEF}\textbf{0.916}
        & \cellcolor[HTML]{EFEFEF}\textbf{0.690} \\
\cmidrule(lr){1-10}
\multirow{5}{*}{GPT-5-mini}
        & Prompting    & 3,323 & 0.790 & 0.473 & 0.632 & 0.883 & 0.840 & 0.741 & 0.617 \\
  &     RepoAgent    & 3,650 & 0.555 & \textbf{1.000} & 0.778 & \underline{0.898} & \underline{0.947} & \underline{0.850} & 0.449 \\
  &     DocAgent     & 3,198 & 0.776 & \underline{0.944} & \underline{0.860} & \underline{0.898} & 0.925 & 0.831 & 0.671 \\
  &     CodeWiki     & 80    & \textbf{0.975} & 0.579 & 0.777 & 0.847 & 0.771 & 0.653 & \underline{0.708} \\
  &     \cellcolor[HTML]{EFEFEF}\textbf{MemDocAgent}
        & \cellcolor[HTML]{EFEFEF}3,323
        & \cellcolor[HTML]{EFEFEF}\underline{0.916}
        & \cellcolor[HTML]{EFEFEF}\textbf{1.000}
        & \cellcolor[HTML]{EFEFEF}\textbf{0.958}
        & \cellcolor[HTML]{EFEFEF}\textbf{0.983}
        & \cellcolor[HTML]{EFEFEF}\textbf{0.968}
        & \cellcolor[HTML]{EFEFEF}\textbf{0.952}
        & \cellcolor[HTML]{EFEFEF}\textbf{0.800} \\
\midrule
\multicolumn{9}{c}{\textit{Closed-Source Baselines}} \\
\midrule
\multicolumn{2}{l}{Claude-Code (\texttt{/init})} & 20  & 1.000* & 0.488 & 0.744 & 0.845 & 0.869 & 0.734 & 0.696 \\
\multicolumn{2}{l}{DeepWiki}  & 404 & 1.000* & 0.840 & 0.920 & 0.815 & 0.935 & 0.762 & 0.755 \\
\bottomrule
\end{tabular}%
}
\end{subtable}

\vspace{1.0em}

\begin{subtable}[t]{\linewidth}
\centering
\caption{Results separated by documentation granularity (component-, module-, and repository-level).}
\label{tab:main_result_levels}
\resizebox{\linewidth}{!}{%
\begin{tabular}{l l ccc ccc ccc}
\toprule
\textbf{Backbone} & \textbf{Method}
& \multicolumn{3}{c}{\textbf{Component-Level}}
& \multicolumn{3}{c}{\textbf{Module-Level}}
& \multicolumn{3}{c}{\textbf{Repository-Level}} \\
\cmidrule(lr){3-5} \cmidrule(lr){6-8} \cmidrule(lr){9-11}
& & Complete. & Truthful. & Helpful.
    & Complete. & Truthful. & Helpful.
    & Complete. & Truthful. & Helpful. \\
\midrule
\multicolumn{11}{c}{\textit{Open-Source Baselines}} \\
\midrule
\multirow{5}{*}{Qwen3-Coder}
  & Prompting    & \underline{0.886} & 0.694 & \underline{0.577} & \underline{0.870} & 0.666 & 0.313 & \underline{0.758} & 0.587 & 0.483 \\
  & RepoAgent    & 0.813 & \underline{0.835} & 0.391 & --    & --    & --    & --    & --    & --    \\
  & DocAgent     & 0.845 & 0.800 & 0.556 & --    & --    & --    & --    & --    & --    \\
  & CodeWiki     & --    & --    & --    & 0.630 & \underline{0.769} & \underline{0.606} & 0.613 & \underline{0.775} & \underline{0.650} \\
  & \cellcolor[HTML]{EFEFEF}\textbf{MemDocAgent}
        & \cellcolor[HTML]{EFEFEF}\textbf{0.987}
        & \cellcolor[HTML]{EFEFEF}\textbf{0.962}
        & \cellcolor[HTML]{EFEFEF}\textbf{0.691}
        & \cellcolor[HTML]{EFEFEF}\textbf{0.975}
        & \cellcolor[HTML]{EFEFEF}\textbf{0.902}
        & \cellcolor[HTML]{EFEFEF}\textbf{0.702}
        & \cellcolor[HTML]{EFEFEF}\textbf{0.975}
        & \cellcolor[HTML]{EFEFEF}\textbf{0.883}
        & \cellcolor[HTML]{EFEFEF}\textbf{0.677} \\
\cmidrule(lr){1-11}
\multirow{5}{*}{GPT-5-mini}
  & Prompting    & 0.631 & 0.761 & 0.627 & 0.626 & \underline{0.766} & 0.626 & 0.640 & 0.696 & 0.598 \\
  & RepoAgent    & 0.778 & \underline{0.850} & 0.449 & --    & --    & --    & --    & --    & --    \\
  & DocAgent     & \underline{0.860} & 0.831 & \underline{0.671} & --    & --    & --    & --    & --    & --    \\
  & CodeWiki     & --    & --    & --    & \underline{0.809} & 0.597 & \underline{0.737} & \underline{0.744} & \underline{0.708} & \underline{0.679} \\
  & \cellcolor[HTML]{EFEFEF}\textbf{MemDocAgent}
        & \cellcolor[HTML]{EFEFEF}\textbf{0.952}
        & \cellcolor[HTML]{EFEFEF}\textbf{0.974}
        & \cellcolor[HTML]{EFEFEF}\textbf{0.750}
        & \cellcolor[HTML]{EFEFEF}\textbf{0.974}
        & \cellcolor[HTML]{EFEFEF}\textbf{0.924}
        & \cellcolor[HTML]{EFEFEF}\textbf{0.858}
        & \cellcolor[HTML]{EFEFEF}\textbf{0.949}
        & \cellcolor[HTML]{EFEFEF}\textbf{0.959}
        & \cellcolor[HTML]{EFEFEF}\textbf{0.791} \\
\bottomrule
\end{tabular}%
}
\end{subtable}

\end{table*}

\subsection{Baselines}
We compare MemDocAgent against four open-source and two closed-source systems. Details are in the Appendix~\ref{B.2}.

\textit{Open-source baselines.}
RepoAgent~\citep{repoagent}, DocAgent~\citep{docagent}, and CodeWiki~\citep{codewiki} are agent-based systems for repository-level documentation generation. We additionally include \textit{Prompting}, a baseline that prompts the LLM to document each target unit independently, with first-hop dependency context.

\textit{Closed-source baselines.}
DeepWiki~\citep{deepwiki} is a commercial system widely used in industry for automated repository-level documentation. We also include Claude-Code~\citep{claudecode}, treating the \texttt{README.md} produced by its \texttt{/init} command as the repository-level documentation.


\subsection{Implementation details}
To evaluate the effectiveness of MemDocAgent on repository-level documentation generation, we select 20 Python repositories from DevEval~\citep{deveval} via stratified sampling, applying three criteria: (i) balanced coverage across the 10 domains in DevEval, (ii) the presence of unit tests to enable information sufficiency evaluation, and (iii) substantial structural depth and dependency complexity at the repository level. Detailed statistics are provided in Appendix~\ref{B.3}.

Our contribution is designing a novel pipeline for repository-level code documentation; the question we aim to answer is therefore which pipeline design yields better documentation. To isolate this effect, we compare all baselines with the same backbone LLMs and hyperparameters and evaluate them under identical settings.
We implement MemDocAgent with two backbone LLMs: Qwen3-Coder-30B-A3B (open-source from HuggingFace) and GPT-5-mini (closed-source from OpenAI). For hyperparameters, each repository is documented with a maximum of 10 steps per sub-trajectory, up to 2 revisions per unit, and a verify threshold of 0.9. 

For evaluation, we use Claude Haiku 4.5 as the LLM judge for truthfulness and helpfulness, and GPT-4o-mini and Claude Haiku 4.5 as code generators for evaluating information sufficiency. To mitigate self-preference bias, we deliberately use evaluator models that differ from any backbone used by MemDocAgent or the baselines. More implementation details are in Appendix~\ref{B.4}.

\section{Experimental results}

\subsection{Comparisons with baselines}~\label{6.1}
\vspace{-2.5em}

\paragraph{Documentation quality comparisons.} 
Table~\ref{tab:main_result} presents a performance comparison between MemDocAgent and existing baselines. Note that the number of generated documents (\#Doc) differs because baselines use different documentation granularities and targets. 
Table~\ref{tab:main_result_overall} shows that MemDocAgent achieves the best overall performance across both open-source backbones. With Qwen3-Coder, it improves completeness, truthfulness, and helpfulness to 0.979, 0.916, and 0.690, outperforming the strongest baselines by 15.9\%, 9.7\%, and 9.9\%, respectively; with GPT-5-mini, it further raises truthfulness and helpfulness to 0.952 and 0.800 while maintaining high completeness at 0.958. Moreover, existing baselines exhibit clear granularity trade-offs. RepoAgent attains high completeness but lower helpfulness, whereas CodeWiki produces concise repository-level documents but misses fine-grained details. In contrast, MemDocAgent generates hierarchical documentation with shared memory, improving coverage, consistency, and utility over both open and closed-source systems.

To ensure a fair comparison within each documentation granularity, we also report results broken down by granularity in Table~\ref{tab:main_result_levels}. The result shows that while existing baselines cover only parts of this hierarchy, MemDocAgent consistently maintains high completeness, truthfulness, and helpfulness across all levels, demonstrating the benefit of long-horizon memory-guided documentation.
Per-repository results, further scalability and generalization analysis are provided in Appendix~\ref{C.1}, ~\ref{C.2}, and ~\ref{C.3}, respectively.



\begin{figure}[t]
\centering
\includegraphics[width=\linewidth]{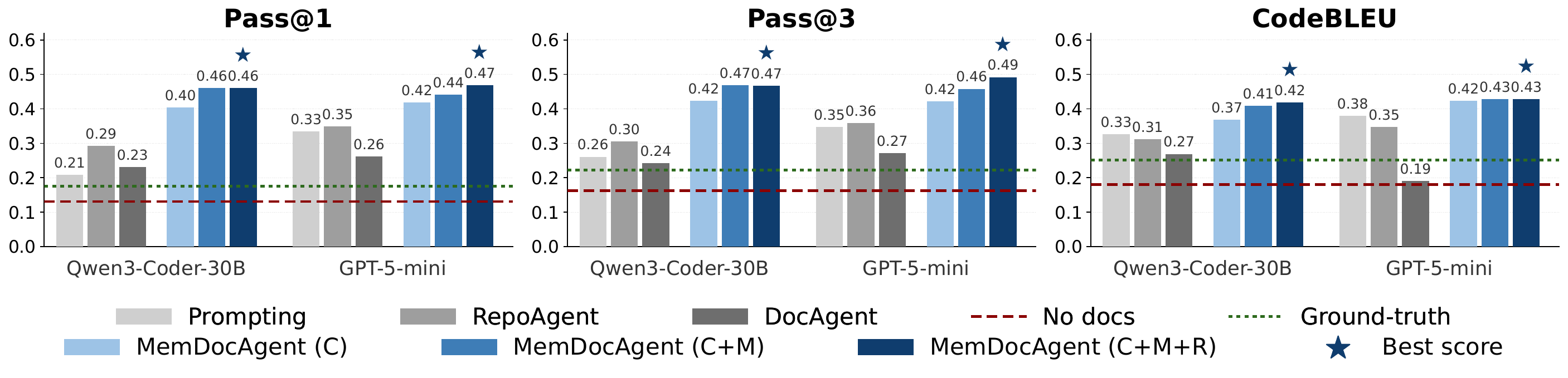}
\caption{Information sufficiency comparison results. Pass@1, @3 and CodeBLEU when regenerating code from each method's documentation alone, averaged over two code generators, with the x-axis denoting the backbone used for each open-source baseline.}
\label{fig:sufficiency_exp}
\end{figure}

\paragraph{Information sufficiency comparisons.}

We measure whether documentation alone provides sufficient information to reproduce the original code by feeding the documentation as the only context to two generation models (GPT-4o-mini and Claude Haiku 4.5) and evaluating Pass@1, Pass@3, and CodeBLEU against ground-truth code. 
To examine whether progressively richer hierarchical context helps the model regenerate code that integrates coherently and executes correctly within the repository, we evaluate three MemDocAgent variants that progressively widen the supplied context: \textit{(C)} the target component document only; \textit{(C+M)} the component with its direct parent module document; and \textit{(C+M+R)} the component with parent module and repository document.
We additionally include two reference settings: \textit{No docs} (signature only) and \textit{Ground-truth} (signature with the human-written component docstring obtained from DevEval~\citep{deveval}).

Figure~\ref{fig:sufficiency_exp} shows the evaluation result (see Appendix~\ref{C.4} for full result.). Across both backbones, every existing baseline surpasses the \textit{No docs} setting, confirming that automatically generated documentation provides useful signal for code reproduction. Among them, MemDocAgent achieves the strongest performance across all settings. Generated documentation can score higher than the \textit{Ground-truth} reference because human-written docstrings are intended as short and concise summaries, whereas the baselines and our method produce richer descriptive documentation.

Compared among three MemDocAgent variants, performance increases monotonically as granularity levels are added, with \textit{(C+M+R)} attaining the best score in every setting. That higher-level documents consistently help indicates that our framework produces cross-level consistent documentation whose repository-wide context and architectural information enable the model to regenerate code that executes correctly within the repository.

\begin{table}[t]
\caption{Ablation study on key components of MemDocAgent. All ablations are conducted under the Qwen3-Coder backbone. Values in parentheses indicate the drop in percentage points relative to ours.}
\label{tab:ablation}
\centering
\resizebox{1\linewidth}{!}{%
\begin{tabular}{l ccc ccc}
\toprule
& \multicolumn{3}{c}{\textbf{Documentation Quality}} & \multicolumn{3}{c}{\textbf{Information Sufficiency}} \\
\cmidrule(lr){2-4} \cmidrule(lr){5-7}
\textbf{Method} & Completeness & Truthfulness & Helpfulness & Pass@1 & Pass@3 & CodeBLEU \\
\midrule
\rowcolor[HTML]{EFEFEF}
\textbf{MemDocAgent} & \textbf{0.979} & \textbf{0.916} & \textbf{0.690} & \textbf{0.461} & \textbf{0.467} & \textbf{0.419} \\
\quad w/o think             & 0.917 \textcolor{red!70!black}{\scriptsize($-$6.2)}  & 0.777 \textcolor{red!70!black}{\scriptsize($-$13.9)} & 0.583 \textcolor{red!70!black}{\scriptsize($-$10.7)} & 0.312 \textcolor{red!70!black}{\scriptsize($-$14.9)} & 0.319 \textcolor{red!70!black}{\scriptsize($-$14.8)} & 0.225 \textcolor{red!70!black}{\scriptsize($-$19.4)} \\
\quad w/o memory system     & 0.903 \textcolor{red!70!black}{\scriptsize($-$7.6)}  & 0.768 \textcolor{red!70!black}{\scriptsize($-$14.8)} & 0.593 \textcolor{red!70!black}{\scriptsize($-$9.7)}  & 0.324 \textcolor{red!70!black}{\scriptsize($-$13.7)} & 0.333 \textcolor{red!70!black}{\scriptsize($-$13.4)} & 0.190 \textcolor{red!70!black}{\scriptsize($-$22.9)} \\
\quad w/o conflict verify   & 0.900 \textcolor{red!70!black}{\scriptsize($-$7.9)}  & 0.781 \textcolor{red!70!black}{\scriptsize($-$13.5)} & 0.592 \textcolor{red!70!black}{\scriptsize($-$9.8)}  & 0.282 \textcolor{red!70!black}{\scriptsize($-$17.9)} & 0.290 \textcolor{red!70!black}{\scriptsize($-$17.7)} & 0.233 \textcolor{red!70!black}{\scriptsize($-$18.6)} \\
\bottomrule
\end{tabular}%
}
\end{table}

\subsection{Ablation study}
We conduct an ablation study to assess the contribution of each proposed component. In Table~\ref{tab:ablation}, we compare with three ablation settings. \textit{w/o think} removes the agent's think step before each action, \textit{w/o memory system} replaces RepoMemory with a stateless cache that is cleared after each documentation unit, and \textit{w/o conflict verify} disables cross-document conflict verification. The result shows that removing any single component causes substantial drops across all metrics, confirming that every component is essential.
The three components affect different aspects of documentation. Together, ReAct-style reasoning, RepoMemory, and conflict verification each contribute a complementary dimension of quality, and only their combination yields MemDocAgent's full hierarchical documentation capability.
We also conduct further ablations: a comparison of memory architectures (Appendix~\ref{C.10}), an analysis of the effectiveness of traversal order (Appendix~\ref{C.6}), prompt-controlled comparisons (Appendix~\ref{C.7}), and an analysis of hyperparameters (Appendix~\ref{C.8}).

\subsection{Efficiency analysis} 

\begin{figure}[t]
    \centering
    \begin{subfigure}[b]{0.3\textwidth}
        \centering
        \includegraphics[width=\textwidth]{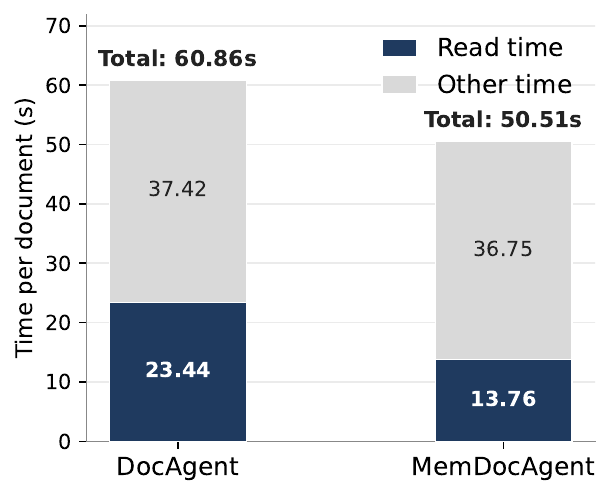}
        \caption{Average read time across 20 repositories.}
        \label{fig:time}
    \end{subfigure}
    \hfill
    \begin{subfigure}[b]{0.68\textwidth}
        \centering
        \includegraphics[width=\textwidth]{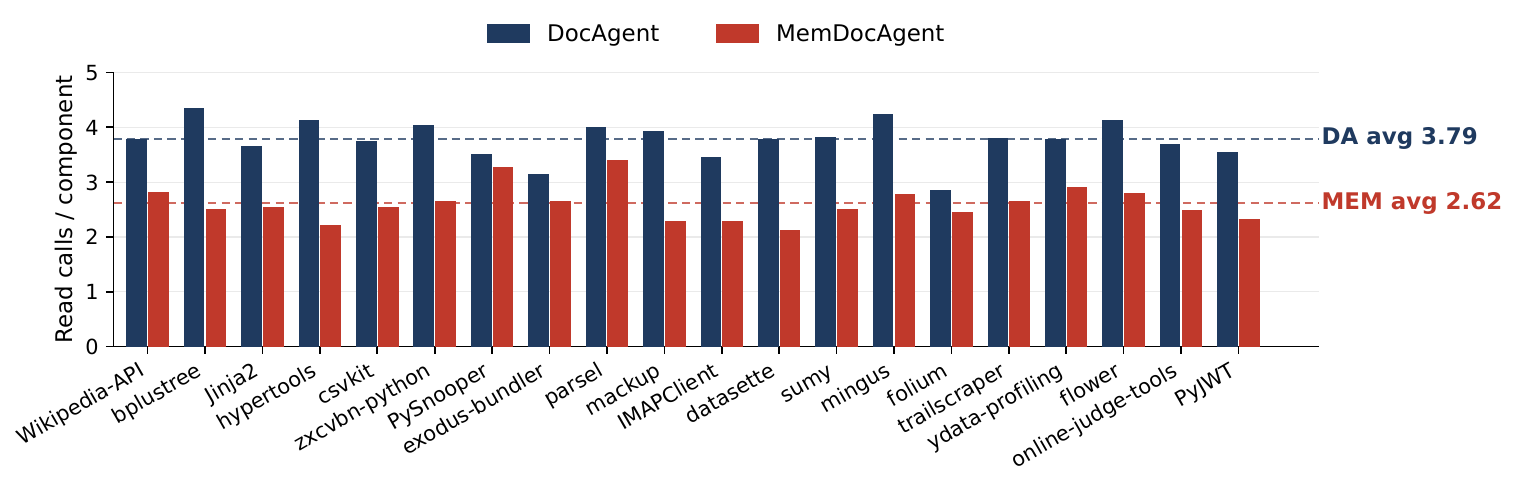}
        \caption{Read calls per component across 20 repositories. Dashed lines mark the overall averages.}
        \label{fig:appen_effi}
    \end{subfigure}
    \caption{Efficiency comparison between DocAgent (previous SOTA) and MemDocAgent.}
    \label{fig:efficiency_combined}
\end{figure}

As shown in Figure~\ref{fig:time}, the effect of eliminating redundant source-file retrieval is consistently observed in retrieval cost. DocAgent performs on average 23.44 seconds, whereas MemDocAgent completes the same task in 13.76 seconds, reducing read time by 41\%. Figure~\ref{fig:appen_effi} confirms that this gain is not driven by a few favorable repositories, as MemDocAgent requires fewer read calls per component in all 20 repositories (3.79 $\to$ 2.62 on average). We attribute this consistent improvement to RepoMemory, which replaces repeated raw-context reads with lookups of verified dependency documents. Further time and cost analyses are reported in Appendix~\ref{C.5}. Together, these results show that MemDocAgent achieves the broadest documentation coverage while remaining more time- and cost-efficient than DocAgent.


\begin{figure}[t]
\centering
\includegraphics[width=1\linewidth]{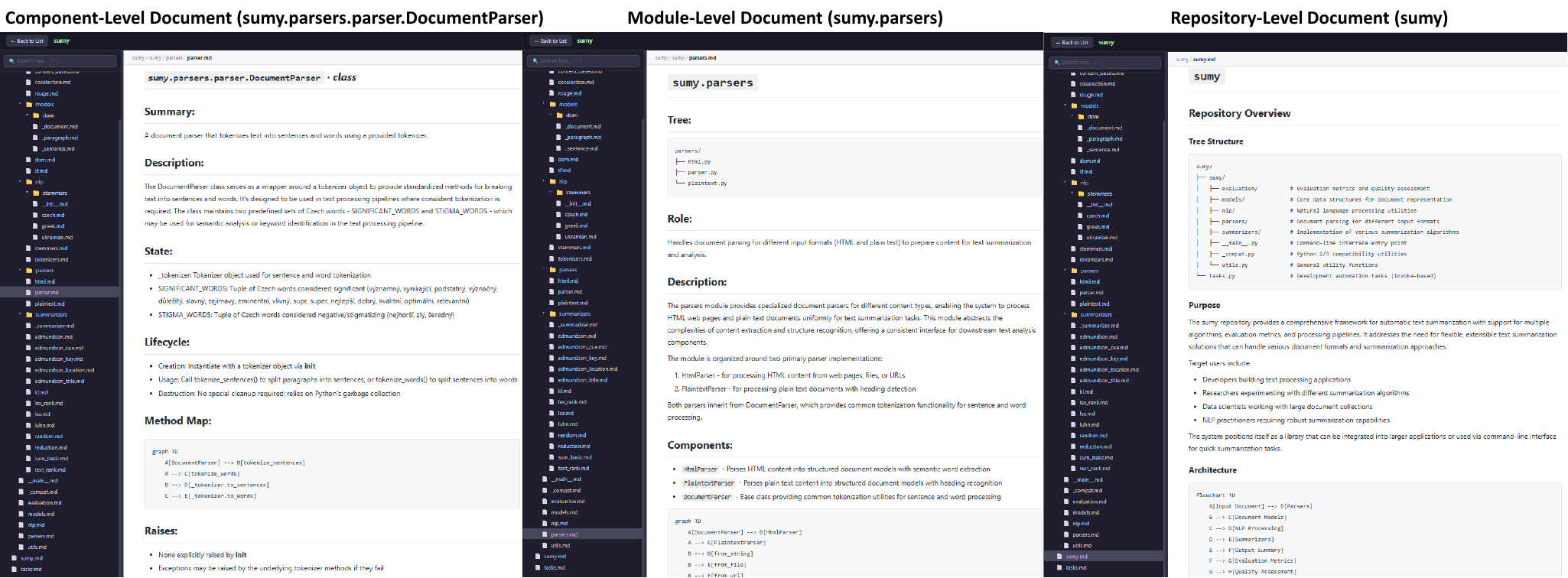}
\caption{Example hierarchical documentation generated by MemDocAgent for the \textit{Sumy} repository, illustrating the three granularity levels (left to right). Each captures granularity-appropriate content, jointly forming a coherent technical from implementation details to high-level architectural patterns.}
\label{fig:case_study}
\end{figure}

\subsection{Case study}
As illustrated in Figure~\ref{fig:case_study}, MemDocAgent generates hierarchical documentation across component-, module-, and repository-levels, ensuring each granularity captures context-appropriate content that ranges from implementation details to high-level architectural patterns. Additional examples are provided in Appendix~\ref{C.9}.

\subsection{Effectiveness of \textsc{Verify} attempts}

\begin{wrapfigure}{r}{0.4\textwidth}
\centering
\vspace{-4em}  
\includegraphics[width=0.4\textwidth]{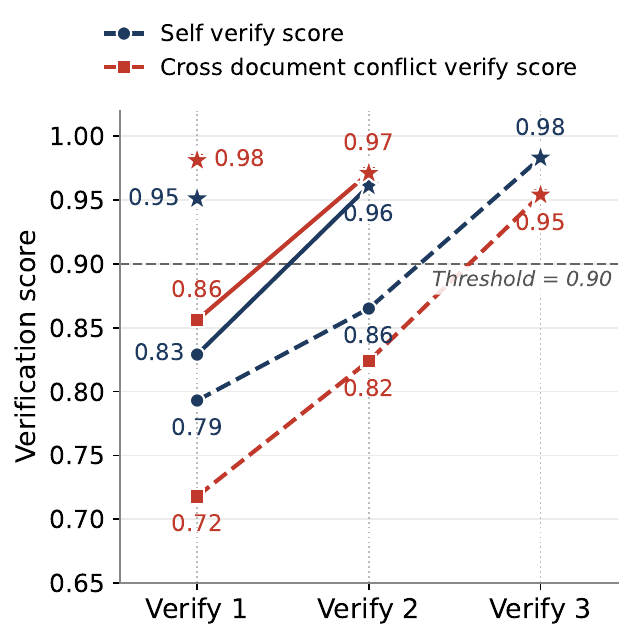}
    \caption{Documentation improvements across \textsc{Verify} attempts.}
    \label{fig:verify}
\vspace{-1em}
\end{wrapfigure}

Figure~\ref{fig:verify} shows how self-evaluation and cross-document conflict scores change over repeated \textsc{Verify} attempts.
63.8\% of components pass on the first attempt with an average score of approximately 0.97, while the remaining 24.4\% and 11.3\% are resolved within two and three attempts, respectively, with documentation quality improving at each additional attempt.
When a component receives a low score at the initial verification, MemDocAgent autonomously refines it, leading to higher self-evaluation and cross-document consistency scores in subsequent attempts and thereby ensuring accurate and consistent documentation. These results demonstrate that the \textsc{Verify} action effectively improves documentation quality through targeted re-verification, applied only to components with detected conflicts or self-evaluation failures.

\section{Conclusion}
\vspace{-0.3em}


We introduced MemDocAgent, a long-horizon agentic framework for repository-level documentation in which a single agent reasons over the entire repository by combining \textit{dependency-aware traversal guiding} and \textit{memory-guided agentic interaction}. MemDocAgent preserves repository-wide context across hierarchical documentation steps, enabling consistent documentation from component-level implementation details to repository-level architecture. Across evaluations, MemDocAgent outperformed existing baselines, improving completeness, truthfulness, helpfulness, and information sufficiency. These results demonstrated the ability of our framework to produce documentation that is more useful for developers and coding agents in real software development workflows.

\section*{Limitation}

Operating over long agentic trajectories introduces several considerations worth discussing. First, long-horizon agentic documentation task requires computational resources because both input context and generated outputs can grow large over the trajectory. To prevent context explosion and excessive runtime, we impose a maximum trajectory length, which can cause rare failures where the agent terminates before completing a sub-task. Second, because the agent depends on the model's reasoning ability, it may occasionally enter a repeated loop without completing the sub-task. Although the overall impact is minimal, with only 0.08\% of components terminating without meeting the verification threshold in our experiments, this limitation should still be considered (Failure analysis in Appendix~\ref{A.4.4}). Future work should explore more efficient trajectory control, adaptive stopping criteria, and stronger recovery mechanisms.
Third, beyond the agent's runtime behavior, our evaluation itself has a scope limitation. We evaluate at the component level information sufficiency because DevEval's human-curated unit tests exist only at the function level, enabling objective, execution-based verification via Pass@k; this is a deliberate choice for metric rigor. We acknowledge it does not directly measure the standalone utility of higher-level documents, but our results offer indirect evidence: regeneration performance increases monotonically as context widens (Figure~\ref{fig:sufficiency_exp}), showing that higher-level documents carry architectural information that materially aids regeneration.

\section*{AI Use Statement}
We used generative AI tools in the following ways:
(i) code development, including implementation of the proposed method, baselines, and experiment scripts; (ii) translation and text editing, for improving clarity and consistency of the manuscript; and (iii) figure and table preparation, including generating plotting scripts and formatting tables.
All AI-generated code was reviewed and tested by the authors. All figures, tables, and reported numbers were checked against the raw experimental outputs, and all references were verified by the authors against their original sources.
The AI tools did not serve as authors. The authors take full responsibility for the entire content of this paper, including all text, claims, figures, tables, references, and artifacts produced with the aid of generative AI.


\bibliographystyle{unsrtnat}

\newpage
\bibliography{reference}
\clearpage
\appendix


\section*{Appendix Contents}
\noindent
\textbf{Appendix~\ref{A}: Additional details about MemDocAgent}
\begin{itemize}
    \item \ref{A.1}\quad Algorithms of dependency-aware traversal guiding
    \item \ref{A.2}\quad System prompts used in MemDocAgent
    \item \ref{A.4}\quad Self-evaluation and cross-document conflict verification: detailed procedure
    \begin{itemize}
        \item \ref{A.4.1}\quad Self-evaluation score
        \item \ref{A.4.2}\quad NLI-based cross-document conflict evaluation score
        \item \ref{A.4.3}\quad Quantitative verification analysis
        \item \ref{A.4.4}\quad Analysis of verification-failure cases
    \end{itemize}
    \item \ref{A.5}\quad Agent trajectory examples on Pysnooper
\end{itemize}

\vspace{0.5em}
\textbf{Appendix~\ref{B}: Additional details about experimental setups}
\begin{itemize}
    \item \ref{B.1}\quad Evaluations
    \item \ref{B.2}\quad Baselines
    \item \ref{B.3}\quad Datasets
    \item \ref{B.4}\quad Implementation details
\end{itemize}

\vspace{0.5em}
\textbf{Appendix~\ref{C}: Additional experimental results}
\begin{itemize}
    \item \ref{C.1}\quad Repository-wise experimental results
    \item \ref{C.2}\quad Scalability analysis
    \item \ref{C.3}\quad Cross-language generalization on large-scale repositories
    \item \ref{C.4}\quad Details of information sufficiency
    \item \ref{C.10}\quad Memory architecture ablation
    \item \ref{C.6}\quad Traversal order ablation
    \item \ref{C.7}\quad Prompt controlled comparisons
    \item \ref{C.8}\quad Hyperparameter sensitivity
    \item \ref{C.5}\quad Efficiency analysis
    \item \ref{C.9}\quad More examples of generated documentation
\end{itemize}

\clearpage


\section{Additional details about MemDocAgent}~\label{A}

\subsection{Algorithms of dependency-aware traversal guiding}\label{A.1}

Algorithm~\ref{alg:build-graph} describes dependency graph construction, and Algorithm~\ref{alg:dfs} presents the topological traversal used for hierarchical generation.

\begin{algorithm}[h]
\caption{Building the unified dependency graph $\mathcal{G}$}
\label{alg:build-graph}
\begin{algorithmic}[1]
\Require Components $\mathcal{C}$ with \textsc{depends\_on}$(c)$ and \textsc{path}$(c)$; repo id $r$
\Ensure  Graph $\mathcal{G}: \mathcal{V} \to 2^{\mathcal{V}}$ over $\mathcal{V} = \mathcal{C} \cup \mathcal{M} \cup \{r\}$, where $u \to v$ means ``$u$ depends on / contains $v$''

\Statex \textit{\# Tier 1: component-level dependency edges}
\For{$c \in \mathcal{C}$}
    \State $\mathcal{G}[c] \gets \textsc{depends\_on}(c)$
    \Comment{call, inheritance, attribute access, import}
\EndFor

\Statex \textit{\# Module hierarchy induced from file paths}
\State Derive $\mathcal{M}$ from $\{\textsc{dirname}(\textsc{path}(c)) : c \in \mathcal{C}\}$ and its prefix closure
\State $\textsc{children}(m) \gets \{\,c \in \mathcal{C} : \textsc{dirname}(\textsc{path}(c)) = m\,\} \cup \{\,m' \in \mathcal{M} : \textsc{dirname}(m') = m\,\}$

\Statex \textit{\# Tier 2: module-level containment edges}
\For{$m \in \mathcal{M}$}
    \State $\mathcal{G}[m] \gets \textsc{children}(m)$
\EndFor

\Statex \textit{\# Tier 3: repository-level containment edges}
\State $\mathcal{G}[r] \gets \{\,v \in \mathcal{M} \cup \mathcal{C} : v \text{ has no parent module}\,\}$
\State \Return $\mathcal{G}$
\end{algorithmic}
\end{algorithm}


\begin{algorithm}[h]
\caption{Calculating dependency-aware traversal order $\pi$}
\label{alg:dfs}
\begin{algorithmic}[1]
\Require Unified graph $\mathcal{G}$ (Alg.~\ref{alg:build-graph})
\Ensure  Order $\pi$ such that (i) every component precedes its dependents, and (ii) every module follows its children, with $r$ last

\State $\mathcal{G} \gets \textsc{CondenseSCC}(\mathcal{G})$ \Comment{Tarjan~\citep{tarjan1972depth}: collapse cycles into super-nodes}
\State $\textit{roots} \gets \{v \in \mathcal{G} : v \text{ has no incoming edge}\}$
\State $\textit{visited} \gets \emptyset$;\quad $\pi \gets [\,]$

\Function{Visit}{$v$} \Comment{iterative post-order; emits $v$ \emph{after} all $\mathcal{G}[v]$}
    \For{$d \in \mathcal{G}[v]$ \textbf{with} $d \notin \textit{visited}$}
        \State \Call{Visit}{$d$}
    \EndFor
    \State $\textit{visited} \mathrel{\cup}= \{v\}$;\quad $\pi.\textsc{append}(v)$
\EndFunction

\For{$v \in \textit{roots} \cup (\mathcal{G} \setminus \textit{visited})$} \Comment{cover isolated nodes too}
    \State \textbf{if} $v \notin \textit{visited}$ \textbf{then} \Call{Visit}{$v$}
\EndFor
\State \Return $\pi$
\end{algorithmic}
\end{algorithm}

\subsection{System prompts used in MemDocAgent}~\label{A.2}

\begin{longtable}{p{\linewidth}}
\caption{Agent's system prompt}
\label{tab:example_1}\\
\toprule
\endfirsthead
\multicolumn{1}{l}{\tablename\ \thetable{} -- \textit{continued from previous page}}\\
\toprule
\endhead
\midrule
\multicolumn{1}{r}{\textit{continued on next page}}\\
\endfoot
\bottomrule
\endlastfoot
\textbf{\texttt{<ROLE>}}\\
You are an expert code documentation worker agent operating on a software repository. Your goal is to generate a high-quality documentation that are both complete and helpful for developers. \\
\textbf{\texttt{</ROLE>}}\\ \\

\texttt{\textbf{<OBJECTIVES>}}\\
Generate documentation sufficient for a developer to reconstruct a functional implementation from scratch using only this documentation.\\
Each level must answer the following questions:\\
- Repository-level : ``What does this system do, and how are its parts organized?''\\
- Module-level : ``What does this module do, and how do its components interact?'' \\
- Component-level: ``How do I implement this function/class/method correctly?''\\
\texttt{\textbf{</OBJECTIVES>}}\\\\

\texttt{\textbf{<DOCUMENTATION\_STRUCTURE>}}\\
Generate documentation following this structure depending on the target level:\\
1. REPO : Repository-level Documentation:\\
\quad - Brief introduction and purpose of the overall system\\
\quad - Architecture overview with diagrams\\
\quad - High-level functionality of each sub-module including references to its documentation file\\
\quad - Link to other module documentation instead of duplicating information\\
\quad - Do not duplicate content covered in MODULE or COMPONENT docs. Focus on the big picture, not implementation details\\\\

2. MODULE: Module-level Documentation:\\
\quad - Explanation of the module's role within the system and its internal design, so a developer can understand *how* its components fit together before reading individual component details\\
\quad - Responsibility and boundaries of the module\\
\quad - List of core components with a one-line description each\\
\quad - Component interaction diagram or call flow (if non-trivial)\\
\quad - Key data structures and state shared across components\\
\quad - Links to component-level documentation for each component\\\\

3. COMPONENT: Component-level Documentation:\\
\quad - Providing enough detail to *reimplement* the function, method, or class correctly --- covering inputs, outputs, behavior, edge cases, and constraints\\
\quad - Summary of what the component does and why it exists (not how it works)\\
\texttt{\textbf{</DOCUMENTATION\_STRUCTURE>}}\\\\

\texttt{\textbf{<WORKFLOW>}}\\
1. You will first receive a sub-task, which includes the type of task (COMPONENT, MODULE, or REPO), the target component/module/repo to document, and other relevant information.\\
2. Analyze the provided code components or module structure, explore the not given dependencies between the components if needed.\\
3. For COMPONENT tasks, generate the documentation for the specific component, and save the documentation in memory with the name of \texttt{`component\_id'}.\\
4. For MODULE tasks, synthesize the documentations of sub-components and generate the module-level documentation, and save the documentation in memory with the name of \texttt{`module\_id'}.\\
5. For REPO tasks, synthesize the documentations of all modules and generate the repository-level documentation, and save the documentation in memory with the name of \texttt{`repo\_id'}.\\
6. For each task, you perform thought-action-observation loops to iteratively improve the documentation until it passes verification, then save the final documentation to memory and return.\\
\quad - At every turn, you MUST follow this structure:\\
\quad\quad Thought: \textlangle your reasoning about what to do next, what information you need\textrangle\\
\quad\quad Action: Choose exactly one action from the list below and provide the necessary input.\\
\quad\quad (Observation will be provided after each action to inform your next step. Do NOT generate this yourself.)\\
\quad - Repeat until all target component is fully documented and verified, then call Finish to save and exit.\\
\quad - The required sequence of actions is typically: Read $\rightarrow$ Write $\rightarrow$ Verify $\rightarrow$ (optional Read or Write) $\rightarrow$ Finish\\
\texttt{\textbf{</WORKFLOW>}}\\\\

\texttt{\textbf{<AVAILABLE\_TOOLS>}}\\
For each turn, first think about what you should do, then take exactly one of the four actions below:\\
- \texttt{READ}: Request specific information about the code component or module structure.\\
- \texttt{WRITE}: Generate documentation for the target component/module/repo based on the provided information and context.\\
- \texttt{VERIFY}: Evaluate the quality of the generated documentation using an external evaluation tool. Use the feedback to iteratively improve the documentation if needed.\\
- \texttt{FINISH}: Save the final documentation to memory and end the task. Only call this after VERIFY passes.\\
\texttt{\textbf{</AVAILABLE\_TOOLS>}}\\\\

\texttt{\textbf{<DETAILED\_TOOL\_USAGE>}}\\
\textbf{1. READ:} If you think more information is needed to generate high-quality documentation of the target component, use this action to request relevant information.\\\\

- During think step, you should analyze the current code and context, and explain what additional information might be needed (if any)\\
- You have access to three types of information sources:\\
\quad 1) Sub-components or sub-modules (from memory):\\
\quad\quad - If the target is a MODULE or REPO, you can request the documentation of its sub-components or sub-modules that have already been documented from memory.\\
\quad\quad - This is the primary source of information for MODULE and REPO tasks, since the module/repo-level documentation should be synthesized based on the already generated sub-component/sub-module documentations.\\
\quad\quad - If you think you need the information of sub-components or sub-modules from memory, use \texttt{<RETRIEVE>YES</RETRIEVE>} tag to request the specific documentations from memory.\\
\quad\quad - If the target is a REPO, you should first read the documentation of all top-level modules and components using \texttt{READ<RETRIEVE>YES</RETRIEVE>}, then decide if you need to read the documentation of lower-level sub-modules based on the gaps you observe in the module-level documentations.\\\\

\quad 2) Internal Codebase Information (from local code repository):\\
\quad\quad For Functions:\\
\quad\quad - Code components called within the function body\\
\quad\quad - Places where this function is called\\
\quad\quad For Methods:\\
\quad\quad - Code components called within the method body\\
\quad\quad - Places where this method is called\\
\quad\quad - The class this method belongs to\\
\quad\quad For Classes:\\
\quad\quad - Code components called in the \texttt{\_\_init\_\_} method\\
\quad\quad - Places where this class is instantiated\\
\quad\quad - Complete class implementation beyond \texttt{\_\_init\_\_}\\\\

\quad 3) External Open Internet retrieval Information:\\
\quad\quad - External Retrieval is extremely expensive. Only request it when understanding an external third-party API or library is essential for accurate documentation, and that information cannot be found within the target codebase.\\
\quad\quad - Use the import statements in \texttt{<IMPORT\_INFORMATION\_IN\_THE\_FILE>} to identify candidates for retrieval. If the component relies on a third-party API (e.g.\ \texttt{from openai import OpenAI}, \texttt{import torch}, \texttt{from stripe import Webhook}), you may request retrieval to understand the external API's expected behavior, parameters, or return types.\\
\quad\quad - Also request retrieval for novel, state-of-the-art algorithms or techniques not self-explanatory from the code alone (e.g.\ NDCG Loss, Cohen's Kappa, specialized metrics).\\
\quad\quad - Each query should be a clear, natural language question targeting specific API behavior or concept.\\\\

- If more information is needed, end your response with a structured request in XML format:\\
\texttt{<RETRIEVE>YES or NO</RETRIEVE>}\\
\texttt{<REQUEST>}\\
\quad \texttt{<INTERNAL>}\\
\quad\quad \texttt{<CLASS>class1,class2</CLASS>}\\
\quad\quad \texttt{<FUNCTION>func1,func2</FUNCTION>}\\
\quad\quad \texttt{<METHOD>self.method1,instance.method2,class.method3</METHOD>}\\
\quad \texttt{</INTERNAL>}\\
\quad \texttt{<RETRIEVAL>}\\
\quad\quad \texttt{<QUERY>query1</QUERY>}\\
\quad \texttt{</RETRIEVAL>}\\
\texttt{</REQUEST>}\\\\

- Important rules for structured request:\\
\quad - If no items exist for a category, use empty tags (e.g., \texttt{<CLASS></CLASS>})\\
\quad - Each external QUERY should be a concise, clear, natural language search query.\\
\quad - Use comma-separated values without spaces for multiple items.\\
\quad - For METHODS, keep dot notation in the same format as the input.\\
\quad - Only first-level calls of the focal code component are accessible. Do not request information on code components that are not directly called by the focal component.\\
\quad - External Retrieval is extremely expensive. Only request external open internet retrieval information if the entity involves a novel, state of the art, recently-proposed algorithms or techniques.\\
\quad\quad (e.g.\ computing a novel loss function (NDCG Loss, Alignment and Uniformity Loss, etc), certain novel metrics (Cohen's Kappa, etc), specialized novel ideas that can not be searched within the target codebase)\\
\quad - Each query should be a clear, natural language question\\
\quad - Only request internal codebase information that you think is necessary for docstring generation task. For some components that is simple and obvious, you do not need any other information for docstring generation.\\
\quad - For module and repository-level documentation, you MUST request \texttt{<RETRIEVE>YES</RETRIEVE>} first before any other information source. The memory-stored summaries are the primary basis for module and repository-level documentation and should always be retrieved at the start.\\
\quad - For module and repository-level documentation, Sources 2) and 3) are supplementary. Since sub-components or sub-modules have already been individually documented in prior steps, their internal or external information has already been gathered. Only request additional internal codebase or external retrieval information if there is a specific gap that cannot be addressed by the memory-stored summaries.\\\\

- Example response:\\
Thought: put your thought here.\\
Action: READ\\
\texttt{<RETRIEVE>NO</RETRIEVE>}\\
\texttt{<REQUEST>}\\
\quad \texttt{<INTERNAL>}\\
\quad\quad \texttt{<CLASS></CLASS>}\\
\quad\quad \texttt{<FUNCTION>execute\_query,connect\_db</FUNCTION>}\\
\quad\quad \texttt{<METHOD>self.process\_data,data\_processor.\_internal\_process</METHOD>}\\
\quad \texttt{</INTERNAL>}\\
\quad \texttt{<RETRIEVAL>}\\
\quad\quad \texttt{<QUERY></QUERY>}\\
\quad \texttt{</RETRIEVAL>}\\
\texttt{</REQUEST>}\\\\

\textbf{2. WRITE:} If you think you have collected sufficient context, use this action and generate the documentation for the target task type.\\\\

- General guidelines for high-quality documentation:\\
\quad - Make documentations actionable and specific: Focus on practical usage.\\
\quad - Use clear, concise language: Avoid jargon unless necessary, use active voice, and be direct and specific.\\
\quad - Type Information: Include precise type hints, note any type constraints, and document generic type parameters.\\
\quad - Context and Integration: Explain component relationships, note any dependencies, and describe side effects.\\
\quad - Follow Google docstring format: Use consistent indentation, maintain clear section separation.\\
- You MUST include the documentation in a \texttt{<DOCUMENTATION>...<\!/DOCUMENTATION>} block including required contents.\\
- The \texttt{<DOCUMENTATION>} block must contain ONLY the documentation text --- section headers and their content.\\
- NEVER include any of the following inside \texttt{<DOCUMENTATION>}:\\
\quad - Python source code (\texttt{def}, \texttt{class}, \texttt{return}, \texttt{import} statements)\\
\quad - Triple-quoted docstrings (triple double-quotes or triple single-quotes)\\
\quad - Code fences (\texttt{```})\\
\quad - The original function/class definition\\
\quad - Any comment syntax (\texttt{\#} or \texttt{//})\\\\

- CORRECT example:\\
Thought: put your thought here.\\
Action: WRITE\\
\texttt{<DOCUMENTATION>}\\
\texttt{\#\# Summary:}\\
Reads a file and returns its content as a list of lines.\\
\texttt{\#\# Description:}\\
Opens the specified file in read mode using the given encoding.\\
\texttt{\#\# Args:}\\
\quad path (str): Absolute or relative path to the target file.\\
\quad encoding (str): Character encoding. Defaults to `utf-8'.\\
\texttt{\#\# Returns:}\\
\quad list[str]: A list of strings, one per line, with trailing newlines removed.\\
\texttt{\#\# Raises:}\\
\quad FileNotFoundError: If the file does not exist.\\
\texttt{</DOCUMENTATION>}\\\\

- WRONG example (do NOT do this):\\
\texttt{<DOCUMENTATION>}\\
\texttt{def read\_lines(path, encoding=`utf-8'):}\\
\quad \texttt{"""Reads a file and returns its content as a list of lines."""}\\
\quad \texttt{return open(path, encoding=encoding).readlines()}\\
\texttt{</DOCUMENTATION>}\\\\

\textbf{3. VERIFY:} After generating a documentation, use this action to self-evaluate the documentation quality along three criteria, each scored from 0.00 to 1.00 (two decimal places).\\\\

- Verification Process:\\
\quad - First read the target task information (source code and related information) as if you're seeing it for the first time.\\
\quad - Read the generated documentation and evaluate each of the following criteria.\\
\quad - Be a harsh, skeptical reviewer. Assume the documentation is flawed until proven otherwise.\\
\quad - A perfect 1.00 is reserved for documentation that cannot be improved in any way.\\\\

- Scoring discipline:\\
\quad - Start each criterion at 0.50 and adjust up or down based on concrete evidence found during review.\\
\quad - Deduct at least 0.05 for each distinct minor flaw (e.g.\ imprecise wording, missing edge case).\\
\quad - Deduct at least 0.10 for each distinct major flaw (e.g.\ wrong type, fabricated parameter, missing parameter entirely).\\
\quad - A score above 0.85 requires an explicit justification in your Thought --- list what the documentation does well.\\
\quad - Never round up. If you are unsure whether a claim is correct, treat it as incorrect.\\
\quad - Avoid anchoring to your own prior output --- evaluate the text as if someone else wrote it.\\\\

- Evaluation criteria:\\
\quad 1) Consistency:\\
\quad\quad - Line-by-line cross-check: for every factual statement in the documentation, locate the corresponding evidence in the source code. If you cannot point to a specific line that supports the claim, deduct points.\\
\quad\quad - Every parameter name, type, and description must exactly match the actual function signature; no parameter is mislabeled, mistyped, or given a fabricated default value.\\
\quad\quad - Described behavior (preconditions, side effects, return values, raised exceptions) must be directly verifiable from the source code --- nothing is fabricated or assumed.\\
\quad\quad - No statement in the documentation may contradict what the code actually does (e.g., wrong return type, incorrect exception condition, swapped parameter roles).\\
\quad\quad - Code examples, if present, must be syntactically valid Python and produce the stated output when executed against the implementation.\\
\quad\quad - Deduct points for: vague hedge words (``typically'', ``usually'', ``may'') used instead of precise statements; claims that go beyond what the source code shows; any detail copied from a different component; invented functionality not present in the code.\\\\

\quad 2) Completeness:\\
\quad\quad - Walk through the source code top-to-bottom and verify that every visible element is documented:\\
\quad\quad\quad * Every parameter: name, type, and purpose. Is any parameter silently omitted?\\
\quad\quad\quad * Return value: type and meaning. Is it documented?\\
\quad\quad\quad * Every exception explicitly raised: type and trigger condition. Is any missing?\\
\quad\quad - Significant edge cases, constraints, and preconditions visible in the source code (e.g.\ if-guards, assertions, boundary checks) must be reflected in the documentation.\\
\quad\quad - For classes: are all public attributes and \texttt{\_\_init\_\_} parameters documented? For modules/repos: are all public components listed with one-line role descriptions?\\
\quad\quad - Deduct points for: any parameter, return value, or raised exception present in the source but absent from the documentation; missing edge-case that is visible in the code; undocumented class attributes or module-level exports.\\\\

\quad 3) Helpfulness:\\
\quad\quad - Read the Summary line in isolation. Does it immediately convey \textit{what} the component does and \textit{why} it exists --- or does it merely restate the function/class name in slightly longer words?\\
\quad\quad - Imagine you are a developer who has never seen this codebase. Can you understand how and when to use this component from the documentation alone, without reading the source code?\\
\quad\quad - Descriptions must explain purpose and typical use cases (WHY / WHEN), not just restate the code logic line-by-line (HOW).\\
\quad\quad - For non-trivial interfaces: are usage examples included? Do they demonstrate realistic, end-to-end scenarios, or are they trivially obvious (e.g.\ just calling the function with no context)?\\
\quad\quad - Deduct points for: summary that paraphrases the function name without adding insight; descriptions that merely narrate the implementation step-by-step; missing examples for complex interfaces; examples that show only the happy path without edge cases.\\\\

- You MUST include a \texttt{<SCORE>} block immediately after the action line with a structured request in XML format:\\
\texttt{<SCORE>}\\
\quad \texttt{<CONSISTENCY>score1</CONSISTENCY>}\\
\quad \texttt{<COMPLETENESS>score2</COMPLETENESS>}\\
\quad \texttt{<HELPFULNESS>score3</HELPFULNESS>}\\
\texttt{</SCORE>}\\\\

- The three scores are combined into a final score to determine whether the documentation meets the verification.\\
- If the observation of the VERIFY action is PASS, proceed to Finish. If the verdict is REVISE, consider whether more context is needed (READ) or whether you can directly rewrite based on the feedback (WRITE).\\
- If the documentation does not pass verification, it may be revised and re-verified up to \texttt{\{max\_revisions\}} times.\\\\

- Example response:\\
Thought: put your thought here.\\
Action: VERIFY\\
\texttt{<SCORE>}\\
\quad \quad \texttt{<CONSISTENCY>0.90</CONSISTENCY>}\\
\quad \quad \texttt{<COMPLETENESS>0.85</COMPLETENESS>}\\
\quad \quad \texttt{<HELPFULNESS>0.80</HELPFULNESS>}\\
\texttt{</SCORE>}\\\\

\textbf{4. FINISH:} If you think the documentation is complete and of high quality, and has passed verification, use this action to save the documentation to memory and end the task. Only call this after VERIFY passes.\\\\

- Example response:\\
Thought: put your thought here.\\
Action: FINISH\\\\
\textbf{\texttt{</DETAILED\_TOOL\_USAGE>}} \\
\end{longtable}

\begin{longtable}{p{\linewidth}}
\caption{Component sub-task injection prompt}
\label{tab:component_user_prompt}\\
\toprule
\endfirsthead
\multicolumn{1}{l}{\tablename\ \thetable{} -- \textit{continued from previous page}}\\
\toprule
\endhead
\midrule
\multicolumn{1}{r}{\textit{continued on next page}}\\
\endfoot
\bottomrule
\endlastfoot

Now, generate comprehensive documentation for the \hlb{function} component named \hlb{bplustree.utils.pairwise} using the provided source code and file path.\\\\
The generated \hlb{function} typed component-level documentation should be included following contents:\\
\hlp{\textbf{\#\# Summary:}\\
\quad - One-line description focusing on WHAT the function does\\
\quad - Avoid repeating the function name\\
\quad - Emphasize the outcome or effect\\

\textbf{\#\# Description:}\\
\quad - List known callers within the codebase and the context in which they call this function: Describe the typical trigger condition or pipeline stage\\
\quad - Explain why this logic is extracted into its own function rather than inlined: Describe the responsibility boundary it enforces\\

\textbf{\#\# Args} (if present)\textbf{:}\\
\quad - name, type, allowed range/values, default value\\
\quad - Note any interdependencies between parameters\\

\textbf{\#\# Returns:}\\
\quad - Explain what the return value represents\\
\quad - Include all possible return values and edge-case return values\\

\textbf{\#\# Raises:}\\
\quad - exception type and exact condition in the code that triggers it\\

\textbf{\#\# Constraints:}\\
\quad - Preconditions: what must be true before calling this function\\
\quad - Postconditions: what is guaranteed to be true after it returns\\

\textbf{\#\# Side Effects:}\\
\quad - Any I/O (files, network, stdout)\\
\quad - External state mutations (global variables, database writes, cache updates)\\
\quad - External service calls\\

\textbf{\#\# Control Flow:}\\
\quad - Mermaid flowchart (flowchart TD) illustrating the main decision branches and loops\\

\textbf{\#\# Examples} (if public and non-trivial)\textbf{:}\\
\quad - Show realistic end-to-end usage including error handling}\\\\

\texttt{<FILE\_PATH>}\\
\hlb{bplustree/utils.py}\\
\texttt{</FILE\_PATH>}\\\\

\texttt{<IMPORT\_INFORMATION\_IN\_THE\_FILE>}\\
\hlb{\texttt{import itertools\\from typing import Iterable}}\\
\texttt{</IMPORT\_INFORMATION\_IN\_THE\_FILE>}\\\\
\texttt{<SOURCE\_CODE>}\\
\hlb{def pairwise(iterable: Iterable):\\
\quad  a, b = itertools.tee(iterable)\\
\quad  next(b, None)\\
\quad  return zip(a, b)}\\
\texttt{</SOURCE\_CODE>}\\

\end{longtable}

\begin{longtable}{p{\linewidth}}
\caption{Module sub-task injection prompt}
\label{tab:module_user_prompt}\\
\toprule
\endfirsthead
\multicolumn{1}{l}{\tablename\ \thetable{} -- \textit{continued from previous page}}\\
\toprule
\endhead
\midrule
\multicolumn{1}{r}{\textit{continued on next page}}\\
\endfoot
\bottomrule
\endlastfoot

Now, generate comprehensive documentation for the \hlb{bplustree} module using the provided module tree.\\\\
The generated module-level documentation should be included following contents:\\
\hlp{\textbf{\#\# Tree:}\\
\quad - Directory/file hierarchy of this module (indented tree format)\\

\textbf{\#\# Role:}\\
\quad - Single-responsibility description: the one thing this module owns\\
\quad - Avoid repeating the module name\\

\textbf{\#\# Description:}\\
\quad - Describe where and when this module is used within the repo: List the primary consumers (other modules or entry points that import it)\\
\quad - Explain why these components are grouped into a separate module: Describe the cohesion principle (shared concept, layer boundary, etc.)\\

\textbf{\#\# Components:}\\
\quad - List all public classes, functions, and constants with their signatures\\
\quad - For each: \textit{one-line} role description\\
\quad - Mermaid dependency graph showing relationships among internal components\\

\textbf{\#\# Public API:}\\
\quad - The interfaces this module exposes to the rest of the repository\\
\quad - For each public symbol: signature, brief description, usage note\\

\textbf{\#\# Dependencies:}\\
\quad - Internal imports (other repo modules) and their purpose\\
\quad - External imports (third-party libraries) and why they are needed\\

\textbf{\#\# Constraints:}\\
\quad - Constraints callers must respect when using this module\\
\quad - Thread-safety, ordering requirements, initialization prerequisites\\}\\
\texttt{<MODULE\_TREE>}\\
\dirtree{%
.1 \colorbox{hlblue}{bplustree/}.
.2 \colorbox{hlblue}{\_\_init\_\_.py}.
.2 \colorbox{hlblue}{const.py}.
.2 \colorbox{hlblue}{entry.py}.
.2 \colorbox{hlblue}{memory.py}.
.2 \colorbox{hlblue}{node.py}.
.2 \colorbox{hlblue}{serializer.py}.
.2 \colorbox{hlblue}{tree.py}.
.2 \colorbox{hlblue}{utils.py}.
}\\
\texttt{</MODULE\_TREE>}\\
\end{longtable}

\begin{longtable}{p{\linewidth}}
\caption{Repository sub-task injection prompt}
\label{tab:repo_user_prompt}\\
\toprule
\endfirsthead
\multicolumn{1}{l}{\tablename\ \thetable{} -- \textit{continued from previous page}}\\
\toprule
\endhead
\midrule
\multicolumn{1}{r}{\textit{continued on next page}}\\
\endfoot
\bottomrule
\endlastfoot

Now, generate comprehensive documentation for the \hlb{bplustree} repository using the provided repository tree.\\\\
The generated repository-level documentation should be included following contents:\\
\hlp{\textbf{\#\# Tree:}\\
\quad - Full top-level directory hierarchy (2--3 levels deep, indented tree format)\\
\quad - Annotate each major directory with its responsibility\\

\textbf{\#\# Purpose:}\\
\quad - What problem this repository solves and why it matters\\
\quad - Target users and the scenarios they use it in\\
\quad - Position in the broader ecosystem (standalone tool, library, service)\\

\textbf{\#\# Architecture:}\\
\quad - End-to-end data flow diagram using Mermaid (flowchart TD or sequence diagram)\\
\quad - Key abstractions and architectural patterns (pipeline, agent loop, plugin, etc.)\\

\textbf{\#\# Entry Points:}\\
\quad - CLI commands, importable APIs, or service endpoints\\
\quad - For each: what it exposes, required arguments, and target audience\\

\textbf{\#\# Core Features:}\\
\quad - List the key capabilities the repo provides\\
\quad - For each feature: one-line description + the implementing module(s)/component(s)\\

\textbf{\#\# Dependencies:}\\
\quad - Key external dependencies and the role each plays\\
\quad - Version constraints or compatibility requirements that affect architecture\\

\textit{Conditional sections:}\\\\

\textbf{\#\# Configuration:}\\
\quad - Config files, environment variables, runtime parameters\\
\quad - Include only if configuration meaningfully affects system behavior\\

\textbf{\#\# Extension Points:}\\
\quad - How to extend the system (plugins, hooks, subclassing, config-driven behavior)\\
\quad - Include only if extensibility is a first-class concern\\}\\

\texttt{<REPO\_TREE>}\\
\dirtree{%
.1 \colorbox{hlblue}{bplustree/}.
.2 \colorbox{hlblue}{bplustree/}.
.2 \colorbox{hlblue}{setup.py}.
}\\
\texttt{</REPO\_TREE>}\\

\end{longtable}

\begin{longtable}{p{\linewidth}}
\caption{Repository-level documentation format}
\label{tab:repo_prompt}\\
\toprule
\endfirsthead
\multicolumn{1}{l}{\tablename\ \thetable{} -- \textit{continued from previous page}}\\
\toprule
\endhead
\midrule
\multicolumn{1}{r}{\textit{continued on next page}}\\
\endfoot
\bottomrule
\endlastfoot

\textbf{\#\# Tree:}\\
\quad - Full top-level directory hierarchy (2--3 levels deep, indented tree format)\\
\quad - Annotate each major directory with its responsibility\\\\

\textbf{\#\# Purpose:}\\
\quad - What problem this repository solves and why it matters\\
\quad - Target users and the scenarios they use it in\\
\quad - Position in the broader ecosystem (standalone tool, library, service)\\\\

\textbf{\#\# Architecture:}\\
\quad - End-to-end data flow diagram using Mermaid (flowchart TD or sequence diagram)\\
\quad - Key abstractions and architectural patterns (pipeline, agent loop, plugin, etc.)\\\\

\textbf{\#\# Entry Points:}\\
\quad - CLI commands, importable APIs, or service endpoints\\
\quad - For each: what it exposes, required arguments, and target audience\\\\

\textbf{\#\# Core Features:}\\
\quad - List the key capabilities the repo provides\\
\quad - For each feature: one-line description + the implementing module(s)/component(s)\\\\

\textbf{\#\# Dependencies:}\\
\quad - Key external dependencies and the role each plays\\
\quad - Version constraints or compatibility requirements that affect architecture\\\\

\textit{Conditional sections:}\\\\

\textbf{\#\# Configuration:}\\
\quad - Config files, environment variables, runtime parameters\\
\quad - Include only if configuration meaningfully affects system behavior\\\\

\textbf{\#\# Extension Points:}\\
\quad - How to extend the system (plugins, hooks, subclassing, config-driven behavior)\\
\quad - Include only if extensibility is a first-class concern\\

\end{longtable}

\begin{longtable}{p{\linewidth}}
\caption{Module-level documentation format}
\label{tab:module_prompt}\\
\toprule
\endfirsthead
\multicolumn{1}{l}{\tablename\ \thetable{} -- \textit{continued from previous page}}\\
\toprule
\endhead
\midrule
\multicolumn{1}{r}{\textit{continued on next page}}\\
\endfoot
\bottomrule
\endlastfoot

\textbf{\#\# Tree:}\\
\quad - Directory/file hierarchy of this module (indented tree format)\\\\

\textbf{\#\# Role:}\\
\quad - Single-responsibility description: the one thing this module owns\\
\quad - Avoid repeating the module name\\\\

\textbf{\#\# Description:}\\
\quad - Describe where and when this module is used within the repo: List the primary consumers (other modules or entry points that import it)\\
\quad - Explain why these components are grouped into a separate module: Describe the cohesion principle (shared concept, layer boundary, etc.)\\\\

\textbf{\#\# Components:}\\
\quad - List all public classes, functions, and constants with their signatures\\
\quad - For each: \textit{one-line} role description\\
\quad - Mermaid dependency graph showing relationships among internal components\\\\

\textbf{\#\# Public API:}\\
\quad - The interfaces this module exposes to the rest of the repository\\
\quad - For each public symbol: signature, brief description, usage note\\\\

\textbf{\#\# Dependencies:}\\
\quad - Internal imports (other repo modules) and their purpose\\
\quad - External imports (third-party libraries) and why they are needed\\\\

\textbf{\#\# Constraints:}\\
\quad - Constraints callers must respect when using this module\\
\quad - Thread-safety, ordering requirements, initialization prerequisites\\

\end{longtable}

\begin{longtable}{p{\linewidth}}
\caption{Component-level (Function) documentation format}
\label{tab:function_prompt}\\
\toprule
\endfirsthead
\multicolumn{1}{l}{\tablename\ \thetable{} -- \textit{continued from previous page}}\\
\toprule
\endhead
\midrule
\multicolumn{1}{r}{\textit{continued on next page}}\\
\endfoot
\bottomrule
\endlastfoot

\textbf{\#\# Summary:}\\
\quad - One-line description focusing on WHAT the function does\\
\quad - Avoid repeating the function name\\
\quad - Emphasize the outcome or effect\\\\

\textbf{\#\# Description:}\\
\quad - List known callers within the codebase and the context in which they call this function: Describe the typical trigger condition or pipeline stage\\
\quad - Explain why this logic is extracted into its own function rather than inlined: Describe the responsibility boundary it enforces\\\\

\textbf{\#\# Args} (if present)\textbf{:}\\
\quad - name, type, allowed range/values, default value\\
\quad - Note any interdependencies between parameters\\\\

\textbf{\#\# Returns:}\\
\quad - Explain what the return value represents\\
\quad - Include all possible return values and edge-case return values\\\\

\textbf{\#\# Raises:}\\
\quad - exception type and exact condition in the code that triggers it\\\\

\textbf{\#\# Constraints:}\\
\quad - Preconditions: what must be true before calling this function\\
\quad - Postconditions: what is guaranteed to be true after it returns\\\\

\textbf{\#\# Side Effects:}\\
\quad - Any I/O (files, network, stdout)\\
\quad - External state mutations (global variables, database writes, cache updates)\\
\quad - External service calls\\\\

\textbf{\#\# Control Flow:}\\
\quad - Mermaid flowchart (flowchart TD) illustrating the main decision branches and loops\\\\

\textbf{\#\# Examples} (if public and non-trivial)\textbf{:}\\
\quad - Show realistic end-to-end usage including error handling\\

\end{longtable}

\begin{longtable}{p{\linewidth}}
\caption{Component-level (Method) documentation format}
\label{tab:method_prompt}\\
\toprule
\endfirsthead
\multicolumn{1}{l}{\tablename\ \thetable{} -- \textit{continued from previous page}}\\
\toprule
\endhead
\midrule
\multicolumn{1}{r}{\textit{continued on next page}}\\
\endfoot
\bottomrule
\endlastfoot

\textbf{\#\# Summary:}\\
\quad - One-line description focusing on WHAT the method does\\
\quad - Avoid repeating the method name\\
\quad - Emphasize the effect on the object's state\\\\

\textbf{\#\# Description:}\\
\quad - List known callers and the context in which they call this method: Describe the lifecycle stage or pipeline step where this method is invoked\\
\quad - Explain why this logic is its own method rather than inlined or placed elsewhere\\\\

\textbf{\#\# Args} (if present)\textbf{:}\\
\quad - name, type, allowed range/values, default value\\\\

\textbf{\#\# Returns:}\\
\quad - type, possible values, and edge-case return values\\\\

\textbf{\#\# Raises:}\\
\quad - exception type and exact condition that triggers it\\\\

\textbf{\#\# State Changes:}\\
\quad - Attributes READ: list \texttt{self.<attr>} fields this method reads\\
\quad - Attributes WRITTEN: list \texttt{self.<attr>} fields this method modifies\\\\

\textbf{\#\# Constraints:}\\
\quad - Preconditions: what must be true about the object/args before calling\\
\quad - Postconditions: what is guaranteed on \texttt{self} or the return value after the call\\\\

\textbf{\#\# Side Effects:}\\
\quad - I/O, external service calls, or mutations to objects outside \texttt{self}\\

\end{longtable}

\begin{longtable}{p{\linewidth}}
\caption{Component-level (Class) documentation format}
\label{tab:class_prompt}\\
\toprule
\endfirsthead
\multicolumn{1}{l}{\tablename\ \thetable{} -- \textit{continued from previous page}}\\
\toprule
\endhead
\midrule
\multicolumn{1}{r}{\textit{continued on next page}}\\
\endfoot
\bottomrule
\endlastfoot

\textbf{\#\# Summary:}\\
\quad - One-line description focusing on WHAT the class represents\\
\quad - Focus on the core purpose or responsibility\\\\

\textbf{\#\# Description:}\\
\quad - Scenarios where this class should be instantiated; known callers/factories that create instances\\
\quad - Motivation for this class as a distinct abstraction; the responsibility boundary it enforces\\\\

\textbf{\#\# State:}\\
\quad - Each attribute: name, type, valid range/values, invariant it participates in\\
\quad - For \texttt{\_\_init\_\_} parameters: note default value and any constraints for the caller\\
\quad - Class invariants: conditions that must always hold between method calls\\\\

\textbf{\#\# Lifecycle:}\\
\quad - Creation: how to instantiate (required args, factory methods)\\
\quad - Usage: which methods are called in what order; any required sequencing\\
\quad - Destruction: cleanup responsibilities (context manager, \texttt{close()}, etc.)\\\\

\textbf{\#\# Method Map:}\\
\quad - Mermaid diagram (flowchart or graph) showing method call dependencies and typical invocation order\\\\

\textbf{\#\# Raises:}\\
\quad - Exceptions raised by \texttt{\_\_init\_\_} and their trigger conditions\\\\

\textbf{\#\# Example:}\\
\quad - Demonstrate creation, typical method sequence, and cleanup\\

\end{longtable}


\subsection{Self-evaluation and cross-document conflict verification: detailed procedure}\label{A.4}

The final verify score combines the self-evaluation score $s_{\text{self}}(\hat{d}_i)$ with the conflict-free fraction $1 - s_{\text{conflict}}(\hat{d}_i)$ via averaging them. $\hat{d}_i$ passes only if the \texttt{verification\_score} exceeds a threshold.

\begin{equation}
    \texttt{verification\_score}(\hat{d}_i) \;=\; \tfrac{1}{2}\bigl(\,s_{\text{self}}(\hat{d}_i) + (1 - s_{\text{conflict}}(\hat{d}_i))\,\bigr) \;\in\; [0, 1].
\end{equation}

\subsubsection{Self-evaluation score} \label{A.4.1}
The agent evaluates the candidate document $\hat{d}_i$ along three criteria: factual consistency, completeness, and helpfulness. For each criterion $c \in \{\text{consistency}, \text{completeness}, \text{helpfulness}\}$, the agent returns a score $s_c(\hat{d}_i) \in [0, 1]$. The overall self-evaluation score is defined as the mean of the three:
\begin{equation}
    s_{\text{self}}(\hat{d}_i) \;=\; \tfrac{1}{3}\bigl(s_{\text{consistency}}(\hat{d}_i) + s_{\text{completeness}}(\hat{d}_i) + s_{\text{helpfulness}}(\hat{d}_i)\bigr).
\end{equation}

\subsubsection{NLI-based cross-document conflict evaluation score} \label{A.4.2}
We verify cross-document conflicts by utilizing existing NLI-based inconsistency detection~\citep{laban2022summac, scire2024fenice, yang-etal-2024-fizz}, which decomposes a candidate text into fine-grained units and checks each unit against a reference source using a pre-trained NLI model. The process consists of the following steps:

\textbf{$\blacktriangleright$ Step 1. Reference set construction.}
We determine the candidate reference set $\mathcal{V}(u_i) \subseteq \text{RepoMemory}$ from the dependency relations of $u_i$, with the granularity of $u_i$ dictating which records are included:
\begin{itemize}\setlength{\itemsep}{0pt}
    \item if $u_i \in \mathcal{C}$ is a component, $\mathcal{V}(u_i)$ contains the component records connected to $u_i$ via the \texttt{depends\_on} relation;
    \item if $u_i \in \mathcal{M}$ is a module, $\mathcal{V}(u_i)$ contains the records of its direct children, the components and sub-modules pointed to by $\mathcal{C}_m$;
    \item if $u_i = r$ is the repository, $\mathcal{V}(u_i)$ contains the records of root modules pointed to by $\mathcal{C}_r$.
\end{itemize}

We then filter $\mathcal{V}(u_i)$ by retaining only the records whose verification score exceeds a threshold ($0.9$), yielding the trusted reference set $\mathcal{D}^{\text{ref}}(u_i)$. If $\mathcal{D}^{\text{ref}}(u_i) = \emptyset$, meaning $u_i$ has no dependencies or none of the candidate references meet the threshold, conflict verification is skipped and we set $s_{\text{conflict}} = 0$.
This structurally prevents a low-quality document from becoming a conflict-checking criterion and propagating errors to dependent components.

\textbf{$\blacktriangleright$ Step2. Hypothesis filtering.}
The verifier first decomposes $\hat{d}_i$ into atomic claims via an LLM (prompt in Table~\ref{tab:claim_extraction_prompt}), and retains only those that explicitly mention a component or module name covered by the reference documents $\mathcal{D}^{\text{ref}}(u_i)$, forming the hypothesis set $\mathcal{H}_i$. This filtering both prevents claims unrelated to the references (e.g., descriptions of $u_i$'s own behavior) from inducing spurious entailment during NLI, and reduces verification cost by reducing NLI calls to reference-relevant claims.

\begin{longtable}{p{\linewidth}}
\caption{Claim extraction prompt for cross-doc conflict verification}
\label{tab:claim_extraction_prompt}\\
\toprule
\endfirsthead
\multicolumn{1}{l}{\tablename\ \thetable{} -- \textit{continued from previous page}}\\
\toprule
\endhead
\midrule
\multicolumn{1}{r}{\textit{continued on next page}}\\
\endfoot
\bottomrule
\endlastfoot
You are a technical documentation analyst.\\
Extract atomic factual claims from the given documentation.\\
Each claim should be a single, verifiable statement about the code's behavior, parameters, return values, side effects, or design.\\
Exclude subjective opinions, vague descriptions, and section headers.\\\\
Documentation:\\
\texttt{\{documentation\}}\\\\
Extract atomic factual claims from this documentation.\\
Return a JSON array of strings. Each string is one claim.\\
Return ONLY the JSON array, no other text.\\\\
Example output:\\
\texttt{["The function returns None when the input is empty",}\\
\texttt{"The timeout parameter defaults to 30 seconds"]}\\
\end{longtable}

\textbf{$\blacktriangleright$ Step 3. Pair-wise NLI verification.}
For each hypothesis $h \in \mathcal{H}_i$ and each reference document $d_j \in \mathcal{D}^{\text{ref}}(u_i)$, we use a pre-trained NLI model\footnote{\texttt{MoritzLaurer/DeBERTa-v3-large-mnli-fever-anli-ling-wanli}.} to test whether $d_j$ entails $h$. A pair $(h, d_j)$ is judged unsupported if the entailment score falls below a threshold $\tau_{\text{nli}}$.
A hypothesis that remains unsupported across every $d_j \in \mathcal{D}^{\text{ref}}(u_i)$ is classified as a conflict, since by construction $h$ shares a component or module with the references (Step 2) yet none of them confirms it. Table~\ref{tab:nli_example} shows representative judgments produced by the NLI model.

\begin{table}[h]
\caption{Examples of NLI judgments used in conflict detection. Premise: ``DataProcessor.process is a static method that returns a list of integers [1, 2, 3]; it takes no arguments.''}
\label{tab:nli_example}
\centering
\resizebox{0.85\linewidth}{!}{%
\begin{tabular}{p{7.5cm}cl}
\toprule
\textbf{Hypothesis (from $\hat{d}_i$)} & \textbf{Label} & \textbf{Reason} \\
\midrule
``process\_data calls DataProcessor.process, which returns a dictionary.'' & contradiction & return type mismatch (list vs.\ dict) \\
``process\_data retrieves data by calling DataProcessor.process, which returns a list.'' & entailment & consistent with premise \\
``process\_data stores the result in \texttt{self.data} for later retrieval.'' & neutral & outside premise's scope \\ \bottomrule
\end{tabular}%
}
\end{table}

\textbf{$\blacktriangleright$ Step 4. Conflict score aggregation.}
We aggregate the per-hypothesis judgments from Step 3 into a document-level conflict score for $\hat{d}_i$.
Let $\mathcal{H}_i^{\text{conf}} \subseteq \mathcal{H}_i$ denote the subset of hypotheses classified as conflict; the conflict score is defined as
\begin{equation}
    s_{\text{conflict}}(\hat{d}_i) \;=\; \frac{|\mathcal{H}_i^{\text{conf}}|}{|\mathcal{H}_i|} \;\in\; [0, 1],
\end{equation}
where larger values indicate greater inconsistency between $\hat{d}_i$ and the reference documents $\mathcal{D}^{\text{ref}}(u_i)$. We set $s_{\text{conflict}}(\hat{d}_i) = 0$ by convention when $\mathcal{H}_i = \emptyset$.

\subsubsection{Quantitative verification analysis}\label{A.4.3}
To quantify the reliability of our verification procedure, three authors independently labeled 50 randomly sampled documents as pass or fail, reaching high agreement (inter-annotator Kappa = 0.85). Treating these human labels as ground truth, we compared them against \textsc{Verify}'s automatic decisions. \textsc{Verify} closely matches human judgments, achieving a precision of 0.967, a recall of 0.935, an F1 of 0.951, and an accuracy of 0.940. These results indicate that \textsc{Verify} reliably distinguishes passing from failing documents, supporting its use as an automatic quality gate in our pipeline.

\subsubsection{Analysis of verification-failure cases}\label{A.4.4}
When a document reaches the revision limit but still falls below the threshold, the document is committed to RepoMemory as its best available version, to preserve full documentation coverage. In our experiments, this occurs for only 0.08\% of all components. 

We manually inspect 29 cases that never passed verification. As a result, None of them contain critical factual errors that could mislead a user. These cases mostly fall into two mechanisms.

\textbf{Case 1 : Unresolved cross-component NLI conflict (18/29)}
The document is factually correct, but a phrasing mismatch with a neighboring component's document keeps the cross-NLI conflict score high enough to block the final score.

\textbf{Case 2 : Self-consistency score plateaus just below threshold (11/29)}
The document is correct, but the self-evaluation score creeps up without quite reaching threshold before the revision budget is exhausted.

Across all cases, no failure stems from a misleading factual error. The common cause is that the strict threshold rejects documents whose cross-NLI penalty or helpfulness sub-score falls marginally short. 
This is also why such documents are never used as trusted references for downstream verification (Appendix~\ref{A.4.2}, Step 1), so they cannot propagate errors even when committed for coverage.


\subsection{Agent trajectory examples on Pysnooper~\cite{rachum2019pysnooper}}~\label{A.5}

\begin{longtable}{p{\linewidth}}
\toprule
\textbf{Subtask \#1 — COMPONENT}: \texttt{misc.generate\_authors.drop\_recurrences} \\
\textit{File}: \texttt{misc/generate\_authors.py} \\
\midrule
\endfirsthead

\multicolumn{1}{l}{\textit{(continued from previous page)}} \\
\toprule
\endhead

\bottomrule
\endfoot

\textbf{Thought1}: Function appears to be a generator that removes duplicates from an iterable while preserving order; no caller context available. \\
\textbf{Action1}: READ\\ \quad \texttt{<RETRIEVE>NO</RETRIEVE>} \\
\textbf{Observation1}: \texttt{No information retrieved.} \\[5pt]

\textbf{Thought2}: I have the function source code and imports available. The function `drop\_recurrences` is straightforward - it removes duplicates from an iterable while preserving order. Since I don't have information about callers or usage context in the broader codebase, I'll document what I can see from the function itself. Let me write the documentation for this function based on the source code analysis.\\
\textbf{Action2}: WRITE<DOCUMENTATION>[..draft]<DOCUMENTATION>\\
\textbf{Observation2}: Draft stored. Proceed to Verify.\\[5pt]

\textbf{Thought3}: Now I need to verify the quality of the documentation I've written. Let me check each criterion\\
\textbf{Action3}: VERIFY\\
\texttt{<SCORE>}\\
\quad \texttt{<CONSISTENCY>0.90</CONSISTENCY>}\\
\quad \texttt{<COMPLETENESS>0.75</COMPLETENESS>}\\
\quad \texttt{<HELPFULNESS>0.85</HELPFULNESS>}\\
\texttt{</SCORE>} \\
\textbf{Observation3}: \textit{Weighted avg} = 0.8333, \textit{Conflict} = 1.00, \textit{Final} = 0.83 $\rightarrow$ REVISE (below threshold 0.90) \\[5pt]

\textbf{Thought4}: The verification shows that completeness needs improvement. Looking at the issues:\\1. The constraint about hashability wasn't properly emphasized - this is a critical limitation of the function\\2. I should be more explicit about what happens with unhashable types (which would cause runtime errors)\\3. The examples could be more comprehensive\\Let me revise the documentation to address these issues.\\
\textbf{Action4}: WRITE<DOCUMENTATION>[..draft]<DOCUMENTATION>\\
\textbf{Observation4}: Draft stored. Proceed to Verify.\\[5pt]

\textbf{Thought5}: Let me verify the revised documentation against the same criteria.\\
\textbf{Action5}: VERIFY\\
\texttt{<SCORE>}\\
\quad \texttt{<CONSISTENCY>0.95</CONSISTENCY>}\\
\quad \texttt{<COMPLETENESS>0.9</COMPLETENESS>}\\
\quad \texttt{<HELPFULNESS>0.9</HELPFULNESS>}\\
\texttt{</SCORE>} \\
\textbf{Observation5}: Verification PASS  (final=0.92, threshold=0.90) \\Proceed to FINISH action. \\[5pt]

\textbf{Thought6}: The documentation has now passed verification with strong scores across all criteria. The function `drop\_recurrences` is well-documented with attention to its key characteristics:\\- It removes duplicates while preserving order\\- It requires hashable elements\\- It raises TypeError for unhashable types\\- It's implemented as a generator for memory efficiency\\- It includes clear examples showing both proper usage and edge cases\\All requirements have been met and the documentation is ready to be saved.\\
\textbf{Action6}: FINISH\\

\midrule
\textbf{Subtask \#31 — COMPONENT}: \texttt{pysnooper.variables.BaseVariable} \\
\textit{File}: \texttt{pysnooper/variables.py} \\
\midrule

\textbf{Thought1}: I need to document the BaseVariable class from pysnooper/variables.py. Let me analyze what I know about this class and what I need to find out. \\
\textbf{Action1}: READ\\
\quad <RETRIEVE>NO</RETRIEVE> \\
\quad\quad<REQUEST>\\
\quad\quad\quad<INTERNAL>\\
\quad\quad\quad\quad<CLASS></CLASS>\\
\quad\quad\quad\quad<FUNCTION>utils.ensure\_tuple,needs\_parentheses</FUNCTION>\\
\quad\quad\quad\quad<METHOD></METHOD>\\\
\quad\quad\quad</INTERNAL>\\
\quad\quad\quad<RETRIEVAL>\\
\quad\quad\quad\quad<QUERY></QUERY>\\
\quad\quad\quad</RETRIEVAL>\\
\quad\quad</REQUEST>\\
\textbf{Observation1}: [Retrieved Context] \\[5pt]

\textbf{Thought2}: I now have information about the utility functions, but I still need to understand what pycompat.ABC is.\\
\textbf{Action2}: READ\\
\quad <RETRIEVE>NO</RETRIEVE> \\
\quad\quad<REQUEST>\\
\quad\quad\quad<INTERNAL>\\
\quad\quad\quad\quad<CLASS></CLASS>\\
\quad\quad\quad\quad<FUNCTION></FUNCTION>\\
\quad\quad\quad\quad<METHOD></METHOD>\\\
\quad\quad\quad</INTERNAL>\\
\quad\quad\quad<RETRIEVAL>\\
\quad\quad\quad\quad<QUERY>pysnooper pycompat module ABC definition</QUERY>\\
\quad\quad\quad</RETRIEVAL>\\
\quad\quad</REQUEST>\\
\textbf{Observation2}: [Retrieved Context] \\[5pt]

\textbf{Thought3}: Let me write the documentation based on what I can definitively determine.\\
\textbf{Action3}: WRITE<DOCUMENTATION>[..draft]<DOCUMENTATION>\\
\textbf{Observation3}: Draft stored. Proceed to Verify.\\[5pt]

\textbf{Thought4}: Now I need to verify this documentation against the source code to ensure consistency, completeness, and helpfulness.\\
\textbf{Action4}: VERIFY\\
\texttt{<SCORE>}\\
\quad \texttt{<CONSISTENCY>0.95</CONSISTENCY>}\\
\quad \texttt{<COMPLETENESS>0.9</COMPLETENESS>}\\
\quad \texttt{<HELPFULNESS>0.85</HELPFULNESS>}\\
\texttt{</SCORE>} \\
\textbf{Observation4}: Verification REVISE  (final=0.82, threshold=0.90)\\Dependency Information No-Conflict : 0.75\\- Conflict with [pysnooper.variables.BaseVariable.items] (support=0.00):\\ YOUR claim : Usage: Call items(frame, normalize=False) to evaluate and process the variable in a given frame context\\DEP states : No direct I/O operations or external service calls\\Fix the above conflicts by aligning your documentation with the dependency's documented behavior.\\Score 0.82 is below threshold 0.90. Revise the documentation.\\Revise the draft (revision 1/2).\\[5pt]

\textbf{Thought5}: I see there's a conflict regarding the usage description. The issue is that I'm claiming the items method ``evaluates and processes the variable in a given frame context'' but the dependency information says there are no direct I/O operations or external service calls. However, looking more carefully at the actual implementation, the items method does perform evaluation (which is a form of processing) and it does take a frame parameter, so my description isn't completely wrong.\\But let me reconsider what's really happening. The method does:\\1. Takes a frame parameter\\2. Evaluates code in that frame context\\3. Calls \_items on the result\\4. Returns the result of \_items\\The conflict seems to be about terminology - ``processes'' might be too vague or misleading. Let me revise to be more precise about what the method actually does.\\
\textbf{Action5}: WRITE<DOCUMENTATION>[..draft]<DOCUMENTATION>\\
\textbf{Observation5}: Draft stored. Proceed to Verify.\\[5pt]

\textbf{Thought6}: Let me verify this revised documentation to make sure it addresses the previous concerns while maintaining accuracy.\\
\textbf{Action6}: VERIFY\\
\texttt{<SCORE>}\\
\quad \texttt{<CONSISTENCY>0.95</CONSISTENCY>}\\
\quad \texttt{<COMPLETENESS>0.95</COMPLETENESS>}\\
\quad \texttt{<HELPFULNESS>0.9</HELPFULNESS>}\\
\texttt{</SCORE>} \\
\textbf{Observation6}: Verification PASS  (final=0.97, threshold=0.90)\\Dependency Information No-Conflict : 1.00\\Proceed to FINISH action. \\[5pt]

\textbf{Thought7}: The documentation has now passed verification with a strong score of 0.97. All criteria have been met.\\
\textbf{Action7}: FINISH\\

\midrule
\textbf{Subtask \#67 — MODULE}: \texttt{pysnooper} \\
\textit{Path}: \texttt{pysnooper/} \\
\midrule

\textbf{Thought1}: I need to generate module-level documentation for the pysnooper module based on the provided directory structure. First, I should retrieve the existing documentation for the sub-modules to understand their individual contributions, then synthesize them into a coherent module-level documentation. Let me start by retrieving the documentation for each submodule that has already been documented in memory.\\
\textbf{Action1}: READ\\
\quad <RETRIEVE>YES</RETRIEVE> \\
\quad\quad<REQUEST>\\
\quad\quad\quad<INTERNAL>\\
\quad\quad\quad\quad<CLASS></CLASS>\\
\quad\quad\quad\quad<FUNCTION></FUNCTION>\\
\quad\quad\quad\quad<METHOD></METHOD>\\\
\quad\quad\quad</INTERNAL>\\
\quad\quad\quad<RETRIEVAL>\\
\quad\quad\quad\quad<QUERY></QUERY>\\
\quad\quad\quad</RETRIEVAL>\\
\quad\quad</REQUEST>\\
\textbf{Observation1}: [Retrieved Context] \\[5pt]

\textbf{Thought2}: Let me write the documentation based on what I can definitively determine.\\
\textbf{Action2}: WRITE<DOCUMENTATION>[..draft]<DOCUMENTATION>\\
\textbf{Observation2}: Draft stored. Proceed to Verify.\\[5pt]

\textbf{Thought3}: Now I need to verify the quality of the generated module-level documentation to ensure it meets the required standards for consistency, completeness, and helpfulness.\\
\textbf{Action3}: VERIFY\\
\texttt{<SCORE>}\\
\quad \texttt{<CONSISTENCY>0.90</CONSISTENCY>}\\
\quad \texttt{<COMPLETENESS>0.85</COMPLETENESS>}\\
\quad \texttt{<HELPFULNESS>0.80</HELPFULNESS>}\\
\texttt{</SCORE>} \\
\textbf{Observation3}: Verification REVISE (final=0.90, threshold=0.90) Dependency Information No-Conflict : 0.94\\
- Conflict with [pysnooper.pycompat.ABC] (support=0.00):\\
YOUR claim : abc: Standard library for abstract base class functionality\\
DEP states : Abstract base class compatibility wrapper for cross-Python-version support.\\
- Conflict with [pysnooper.utils.WritableStream] (support=0.00):\\
YOUR claim : Output destinations must be one of the supported types (None, str, PathLike, callable, or WritableStream)\\
DEP states : An abstract base class defining the interface for writable streams that must implement a write method.\\
Fix the above conflicts by aligning your documentation with the dependency's documented behavior.\\Score 0.90 is below threshold 0.90. Revise the documentation.\\Revise the draft (revision 1/2).\\[5pt]

\textbf{Thought4}: \\
\textbf{Action4}: WRITE<DOCUMENTATION>[..draft]<DOCUMENTATION>\\
\textbf{Observation4}: Draft stored. Proceed to Verify.\\[5pt]

\textbf{Thought5}: Let me verify once more to see if this latest version resolves the remaining conflicts.\\
\textbf{Action5}: VERIFY\\
\texttt{<SCORE>}\\
\quad \texttt{<CONSISTENCY>0.95</CONSISTENCY>}\\
\quad \texttt{<COMPLETENESS>0.95</COMPLETENESS>}\\
\quad \texttt{<HELPFULNESS>0.9</HELPFULNESS>}\\
\texttt{</SCORE>} \\
\textbf{Observation5}: Verification PASS  (final=0.92, threshold=0.90)\\
Dependency Information No-Conflict : 0.94. Proceed to FINISH action.\\[5pt]

\textbf{Thought6}: All criteria have been met.\\
\textbf{Action6}: FINISH\\

\midrule
\textbf{\#68 — REPO}: \texttt{PySnooper} \\
\textit{Path}: \texttt{/home/sybae/project/Data/Source\_Code\_Original/Software-Development/PySnooper} \\
\midrule

\textbf{Thought1}: I need to generate comprehensive repository-level documentation for the PySnooper repository.\\
\textbf{Action1}: READ\\
\quad <RETRIEVE>YES</RETRIEVE> \\
\quad\quad<REQUEST>\\
\quad\quad\quad<INTERNAL>\\
\quad\quad\quad\quad<CLASS></CLASS>\\
\quad\quad\quad\quad<FUNCTION></FUNCTION>\\
\quad\quad\quad\quad<METHOD></METHOD>\\\
\quad\quad\quad</INTERNAL>\\
\quad\quad\quad<RETRIEVAL>\\
\quad\quad\quad\quad<QUERY></QUERY>\\
\quad\quad\quad</RETRIEVAL>\\
\quad\quad</REQUEST>\\
\textbf{Observation1}: [Retrieved Context] \\[5pt]

\textbf{Thought2}: I now have the documentation for the main pysnooper module. I need to also check the misc directory to understand what's in there, as it might contain important information like README or other documentation files. Let me also get information about setup.py to understand the package configuration.\\
\textbf{Action2}: READ\\
\quad <RETRIEVE>YES</RETRIEVE> \\
\quad\quad<REQUEST>\\
\quad\quad\quad<INTERNAL>\\
\quad\quad\quad\quad<CLASS></CLASS>\\
\quad\quad\quad\quad<FUNCTION>read\_file</FUNCTION>\\
\quad\quad\quad\quad<METHOD></METHOD>\\\
\quad\quad\quad</INTERNAL>\\
\quad\quad\quad<RETRIEVAL>\\
\quad\quad\quad\quad<QUERY></QUERY>\\
\quad\quad\quad</RETRIEVAL>\\
\quad\quad</REQUEST>\\
\textbf{Observation2}: [Retrieved Context] \\[5pt]

\textbf{Thought3}: Let me write the documentation based on what I can definitively determine.\\
\textbf{Action3}: WRITE<DOCUMENTATION>[..draft]<DOCUMENTATION>\\
\textbf{Observation3}: Draft stored. Proceed to Verify.\\[5pt]

\textbf{Thought3}: Now I need to verify this repository-level documentation to ensure it's accurate and complete.\\
\textbf{Action3}: VERIFY\\
\texttt{<SCORE>}\\
\quad \texttt{<CONSISTENCY>0.85</CONSISTENCY>}\\
\quad \texttt{<COMPLETENESS>0.80</COMPLETENESS>}\\
\quad \texttt{<HELPFULNESS>0.85</HELPFULNESS>}\\
\texttt{</SCORE>} \\
\textbf{Observation3}: Verification REVISE  (final=0.81, threshold=0.90)\\Dependency Information No-Conflict : 0.78\\
- Conflict with [pysnooper] (support=0.00):\\
YOUR claim : CLI Commands: None - PySnooper is a library, not a command-line tool\\
DEP states : Provides comprehensive code tracing and variable inspection capabilities for debugging Python applications\\
- Conflict with [pysnooper] (support=0.00):\\
YOUR claim : PySnooper operates with minimal configuration:\\
DEP states : Provides comprehensive code tracing and variable inspection capabilities for debugging Python applications\\
Fix the above conflicts by aligning your documentation with the dependency's documented behavior.\\
Score 0.81 is below threshold 0.90. Revise the documentation. Revise the draft (revision 1/2).\\[5pt]

\textbf{Thought4}: I now have the documentation for the main pysnooper module. I need to also check the misc directory to understand what's in there, as it might contain important information like README or other documentation files. Let me also get information about setup.py to understand the package configuration.\\
\textbf{Action4}: READ\\
\quad <RETRIEVE>YES</RETRIEVE> \\
\quad\quad<REQUEST>\\
\quad\quad\quad<INTERNAL>\\
\quad\quad\quad\quad<CLASS>Tracer</CLASS>\\
\quad\quad\quad\quad<FUNCTION>snoop</FUNCTION>\\
\quad\quad\quad\quad<METHOD>\_call\_\_, \_\_enter\_\_, \_\_exit\_\_</METHOD>\\\
\quad\quad\quad</INTERNAL>\\
\quad\quad\quad<RETRIEVAL>\\
\quad\quad\quad\quad<QUERY></QUERY>\\
\quad\quad\quad</RETRIEVAL>\\
\quad\quad</REQUEST>\\
\textbf{Observation4}: [Retrieved Context] \\[5pt]

\textbf{Thought5}: \\
\textbf{Action5}: WRITE<DOCUMENTATION>[..draft]<DOCUMENTATION>\\
\textbf{Observation5}: Draft stored. Proceed to Verify.\\[5pt]

\textbf{Thought6}: Let me verify once more to see if this latest version resolves the remaining conflicts.\\
\textbf{Action6}: VERIFY\\
\texttt{<SCORE>}\\
\quad \texttt{<CONSISTENCY>0.95</CONSISTENCY>}\\
\quad \texttt{<COMPLETENESS>0.9</COMPLETENESS>}\\
\quad \texttt{<HELPFULNESS>0.9</HELPFULNESS>}\\
\texttt{</SCORE>} \\
\textbf{Observation6}: Verification PASS  (final=0.90, threshold=0.90)\\
Dependency Information No-Conflict : 0.88. Proceed to FINISH action.\\[5pt]

\textbf{Thought6}: All criteria have been met.\\
\textbf{Action6}: FINISH\\
\end{longtable}


\section{Additional details about experimental setups}~\label{B}

\subsection{Evaluations}~\label{B.1}

\subsubsection{Completeness} 
Completeness measures the extent to which the generated documentation adheres to standard structural conventions and includes essential code entities expected for a given target. It is computed as the average of two sub-scores: 

\begin{equation}
\text{Completeness} = \frac{\text{Section Presence Score} + \text{Entity Coverage Score}}{2}
\in [0, 1]
\end{equation}

\textbf{Section Presence Score:} For each documentation, we measure the proportion of required sections that are present, using rule-based pattern matching over section headers. The set of required and conditional sections differs per granularity level, reflecting the structural conventions adopted by prior systems (DocAgent, RepoAgent, CodeWiki) and our own framework. Conditional sections are not penalized when absent if they are not applicable to the target (e.g., a function with no arguments is not required to contain an Arguments section).
The required and conditional sections per granularity are summarized in Table~\ref{tab:section-schema}.

\begin{table}[t]
\caption{Required and conditional sections for documentation in each granularity.}
\label{tab:section-schema}
\centering
\resizebox{0.9\linewidth}{!}{%
\begin{tabular}{p{0.1\linewidth} p{0.42\linewidth} p{0.35\linewidth}}
\toprule
\textbf{Granularity} & \textbf{Required Sections} & \textbf{Conditional Sections} \\
\midrule
Function or Method & Summary, Description, Returns & Arguments, Exceptions, Side Effects, Control Flow, Usage Examples \\ \midrule
Class & Summary, Description, Returns & Arguments, Exceptions, Usage Examples, Side Effects \\ \midrule
Module & Module Structure (Tree), Role, Description, Components, Public API, Dependencies & - \\ \midrule
Repository & Repository Structure (Tree), Purpose, Architecture, Entry Points, Core Features, Dependencies & Configuration, Extension Points \\
\bottomrule
\end{tabular}%
}
\end{table}

The Section Presence score is computed as:
\begin{equation}
\text{Section Presence Score} = \frac{|\text{required sections present}|}{|\text{required sections}|}
\end{equation}

\textbf{Entity Coverage Score:} We extract the set of core entities from the target source code via abstract syntax tree (AST) analysis, and measure the proportion of these entities that are referenced in the documentation via regular-expression matching. The core entity set is defined per granularity:
\begin{itemize}
    \item Component-level: Function, class, method, and variable names that appear within the target source code.
    \item Module-level: Names of the child components contained in the module.
    \item Repository-level: Names of the top-level modules together with the component names along their immediate paths.
\end{itemize}

The Entity Coverage score is computed as:
\begin{equation}
\text{Entity Coverage Score} = \frac{|\text{mentioned entities}|}{|\text{total entities}|}
\end{equation}

\subsubsection{Helpfulness}
We assess the semantic quality and practical utility of each documentation through an LLM-as-judge protocol, following the design adopted in DocAgent~\citep{docagent}. The judge rates each required section on a 5-point Likert scale using a section-specific rubric, and the normalized section scores are averaged into a single Helpfulness score:

\begin{equation}
\text{Helpfulness} = \frac{\text{Score} - 1}{4} \in [0, 1]
\end{equation}

We use \texttt{Claude Haiku 4.5 (Anthropic API, temperature = 0.0, max tokens = 256)} as the judge model. The section-specific rubrics in each granularity (component, module, and repository) are described in Table~\ref{tab:component-eval-criteria}, Table~\ref{tab:module-eval-criteria}, and Table~\ref{tab:repo-eval-criteria}, respectively.

\begin{table}[h]
\caption{Component-level helpfulness evaluation criteria. Args and Returns apply to functions and methods only; Summary and Description apply to all component types (function, method, class).}
\label{tab:component-eval-criteria}
\centering
\resizebox{0.8\linewidth}{!}{%
\begin{tabular}{l c p{9.2cm}}
\toprule
\textbf{Aspect} & \textbf{Score} & \textbf{Criterion} \\
\midrule
\multirow{5}{*}{Summary}
 & 1 & Merely restates the component name; no additional value. \\
 & 2 & Adds one minor detail but still vague about purpose. \\
 & 3 & Some context about what it does, but \textsc{why}/\textsc{when} is unclear. \\
 & 4 & Clearly conveys \textsc{what} and \textsc{why} in practical terms. \\
 & 5 & Perfectly balances conciseness and purpose; developer immediately
       understands when/why to use this. \\
\midrule
\multirow{5}{*}{Description}
 & 1 & Only repeats what the code does line-by-line; no context. \\
 & 2 & Mentions callers or motivation superficially. \\
 & 3 & One of \textsc{where}/\textsc{when} or \textsc{why} is well-explained
       but the other is missing. \\
 & 4 & Both addressed clearly. \\
 & 5 & Both addressed with specific detail; a new developer would understand
       the design rationale immediately. \\
\midrule
\multirow{5}{*}{Args}
 & 1 & Just restates parameter types already visible from signature. \\
 & 2 & Adds minimal description; constraints/allowed values missing. \\
 & 3 & Describes purpose of most params but lacks constraints or edge cases. \\
 & 4 & Each param has purpose, type, and constraints; interdependencies noted. \\
 & 5 & Comprehensive: every param's purpose, constraints, interdependencies,
       and edge values documented. \\
\midrule
\multirow{5}{*}{Returns}
 & 1 & Only restates the return type annotation. \\
 & 2 & Brief description; edge-case return values missing. \\
 & 3 & Normal return value explained; some edge cases missing. \\
 & 4 & All return values and their meanings explained, including edge cases. \\
 & 5 & Every possible return value, meaning, and edge condition documented. \\
\bottomrule
\end{tabular}%
}
\end{table}

\begin{table}[h]
\caption{Module-level helpfulness evaluation criteria.}
\label{tab:module-eval-criteria}
\centering
\resizebox{0.8\linewidth}{!}{%
\begin{tabular}{l c p{9.2cm}}
\toprule
\textbf{Aspect} & \textbf{Score} & \textbf{Criterion} \\
\midrule
\multirow{5}{*}{Role}
 & 1 & Just restates the module name or directory path. \\
 & 2 & Vague purpose; no distinction from other modules. \\
 & 3 & Clear responsibility described, but scope boundaries unclear. \\
 & 4 & Precisely describes the one thing this module owns, clearly distinct
       from neighboring modules. \\
 & 5 & Single-responsibility stated so clearly that a developer can immediately
       decide whether to use this module. \\
\midrule
\multirow{5}{*}{Description}
 & 1 & Neither primary consumers nor cohesion principle addressed. \\
 & 2 & One is mentioned superficially. \\
 & 3 & One of consumers or cohesion is well-explained. \\
 & 4 & Both addressed clearly. \\
 & 5 & Both addressed with specific detail; developer understands the
       architectural motivation. \\
\midrule
\multirow{5}{*}{Public API}
 & 1 & Just lists names with no descriptions. \\
 & 2 & Minimal descriptions; usage notes missing. \\
 & 3 & Most symbols described; some usage notes present. \\
 & 4 & All public symbols with signatures, brief descriptions, and usage notes. \\
 & 5 & Complete API reference with guidance on which symbol to use for which task. \\
\bottomrule
\end{tabular}%
}
\end{table}

\begin{table}[h]
\caption{Repository-level helpfulness evaluation criteria.}
\label{tab:repo-eval-criteria}
\centering
\resizebox{0.8\linewidth}{!}{%
\begin{tabular}{l c p{9.2cm}}
\toprule
\textbf{Aspect} & \textbf{Score} & \textbf{Criterion} \\
\midrule
\multirow{5}{*}{Purpose}
 & 1 & Vague; could describe any software project. \\
 & 2 & Names the domain but not the specific problem solved. \\
 & 3 & Clear problem statement, but target users or ecosystem position unclear. \\
 & 4 & Problem, target users, and position in ecosystem all described. \\
 & 5 & A new contributor can immediately understand what, who, and why. \\
\midrule
\multirow{5}{*}{Architecture}
 & 1 & No diagram or abstract description only. \\
 & 2 & High-level description without data flow. \\
 & 3 & Data flow described in text; diagram absent or incomplete. \\
 & 4 & End-to-end data flow diagram present with key abstractions identified. \\
 & 5 & Complete architecture diagram with flow + abstractions + key design patterns. \\
\midrule
\multirow{5}{*}{Core Features}
 & 1 & Just a list of module names without descriptions. \\
 & 2 & Feature descriptions are vague. \\
 & 3 & Most features described with implementing modules. \\
 & 4 & Each feature has a clear one-line description and implementing module(s). \\
 & 5 & Feature-module mapping complete and clear; a developer can navigate to
       any feature immediately. \\
\bottomrule
\end{tabular}%
}
\end{table}

\subsubsection{Truthfulness}
Truthfulness measures whether the content of the generated documentation faithfully reflects the actual source code. It is decomposed into two sub-scores: \textit{Consistency}, which measures factual correctness with respect to the code, and \textit{Relevance}, which measures the proportion of code-relevant content within the documentation. 
%
We adopt a fine-grained documentation evaluation framework that decomposes generated documentation and evaluates each segment using predefined evaluation criteria~\citep{bae-etal-2026-referee, min-etal-2023-factscore} to compute above two scores.

\textbf{Step1: Claim extraction.} From each documentation, we extract natural-language sentences while excluding non-assertive elements (section headers, directory trees, code blocks or diagrams). The remaining text is decomposed into atomic claims following the FactScore~\citep{min-etal-2023-factscore}. Claim extraction is performed by \texttt{GPT-4o-mini} (\texttt{temperature = 0.0, max tokens = 1024}).

\textbf{Step2: Reference context retrieval.} For each documentation target, we collect the relevant source code and its dependency context as the reference, using a granularity-specific strategy:
\begin{itemize}
    \item Component: The target source code together with the source of its dependency components.
    \item Module: The source code of all components contained in the module.
    \item Repository: A name-index–based dynamic retrieval, where component and module names appearing in each claim are matched against the repository index to retrieve the relevant files on demand.
\end{itemize}

\textbf{Step3: Claim-level fine-grained evaluation.} Each atomic claim is evaluated as one of three categories: consistent, inconsistent, or irrelevant, conditioned on the target source code with related information. Evaluation is performed by \texttt{Claude Haiku 4.5 (temperature = 0.0, max tokens = 16)}, separately from the extraction model, so that the extraction and judgment LLMs are decoupled.

\textbf{Step4: Calculating truthfulness.} The \textit{Consistency} score therefore measures the proportion of verifiable claims (excluding off-topic content) that are factually supported by the source code, while the \textit{Relevance} score measures the proportion of the documentation that is on-topic and amenable to verification at all. The two sub-scores are then combined multiplicatively:
\begin{equation}
\text{Consistency} = \frac{\text{Consistent}}{\text{Consistent} + \text{Inconsistent}}
\end{equation}
\begin{equation}
\text{Relevance} = 1-\frac{\text{Irrelevant}}{\text{Consistent} + \text{Inconsistent} + \text{Irrelevant}}
\end{equation}

\begin{equation}
\text{Truthfulness} = \text{Consistency} \times \text{Relevance} \in [0, 1]
\end{equation}

\subsubsection{Information sufficiency}

\begin{figure}[h]
\centering
\includegraphics[width=1\linewidth]{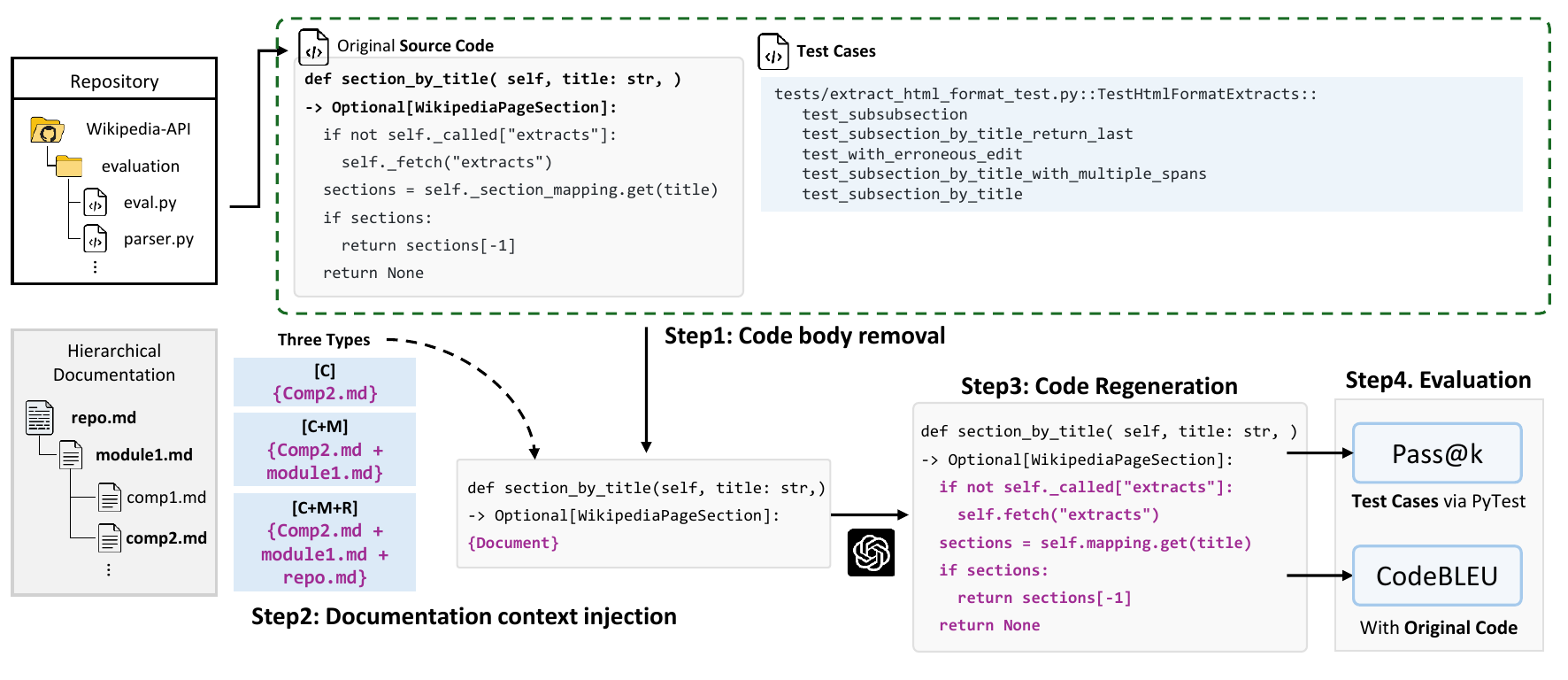}
\caption{Overview of evaluating information sufficiency.}
\vspace{-0.4em}
\label{fig:appen_sufficiency}
\end{figure}

Figure~\ref{fig:appen_sufficiency} illustrates the overall pipeline of our information sufficiency evaluation. DevEval~\citep{deveval} provides human-curated unit tests aligned with reference code, enabling functional verification at the function level. From this resource, we collect 564 test cases targeting 312 component-level functions across our 20 benchmark repositories, and use them as the evaluation set. 
The evaluation proceeds in four stages. 

\textbf{Step1: Code body removal.} We remove the body of each target component-level function, retaining only its signature. 

\textbf{Step2: Documentation context injection.} A code-generation LLM is provided with the function signature together with documentation context drawn from the hierarchical documentation tree, and we consider three progressively richer configurations. \textbf{C} provides only the component's own documentation $d^{(c)}$; \textbf{C+M} additionally includes the parent module documentation $d^{(m)}$; and \textbf{C+M+R} further adds the repository-level documentation $d^{(r)}$, yielding the full hierarchical context $\{d^{(c)}, d^{(m)}, d^{(r)}\}$. This design isolates the contribution of each documentation level to the model's ability to reconstruct the original function.

\textbf{Step3: Function regeneration.} Given the injected context, the LLM regenerates the body of the function. 

\textbf{Step4: Evaluation.} The regenerated function is reinserted into its original location in the repository, after which we (i) execute the corresponding unit tests to verify functional correctness (Pass@k) and (ii) measure the structural similarity between the regenerated and the original source code (CodeBLEU).

\subsection{Baselines}~\label{B.2}
We compare MemDocAgent against six baselines spanning two categories: open-source documentation systems whose internal pipelines are accessible for measurement, and closed-source commercial systems that operate as black boxes.

\paragraph{Open-source baselines.}

\begin{itemize}
    \item \textbf{Prompting}: A non-agentic baseline that prompts an LLM once with the target component and its first-hop dependencies as context, generating documentation in a single forward pass without iterative refinement or cross-document verification. We apply this baseline independently at the component, module, and repository levels.
    \item \textbf{RepoAgent}~\citep{repoagent}: An agentic system that generates function and class-level docstrings by traversing the repository's dependency graph constructed via static analysis. While effective for component-level cataloging, it does not synthesize parent module or repository-level documentation from the underlying components.

    \item \textbf{DocAgent}~\citep{docagent}: A multi-agent framework consisting of Reader, Searcher, Writer, Verifier, and Orchestrator agents. It traverses the code dependency graph in topological order and generates component-level docstrings in a bottom-up manner. Since each component is processed in an isolated context, cross-document consistency with previously generated documents is not explicitly enforced.
    
    \item \textbf{CodeWiki}~\citep{codewiki}: A wiki-style documentation system that produces hierarchical pages at the module and repository levels by clustering related components into structured wiki articles. It targets high-level overviews and does not generate fine-grained component-level documentation.
    
\end{itemize}

\paragraph{Closed-source baselines.}

\begin{itemize}
    \item \textbf{DeepWiki}~\citep{deepwiki}: A commercial system widely used in industry for automated repository-level documentation. As the system's internal pipeline is not publicly accessible, we evaluate only its final output.
    \item \textbf{Claude Code}~\citep{claudecode}: A command-line coding assistant. We treat the \texttt{README.md} produced by its \texttt{/init} command as the repository-level documentation.
\end{itemize}

\subsection{Datasets}~\label{B.3}
Table~\ref{tab:dataset} summarizes the statistics of the \textbf{20 repositories} used in our evaluation. The 20 repositories in our evaluation span \textbf{10 distinct domains} (Communication, Internet, Utilities, etc.), differing substantially in code style and repository structure.

\begin{table}[h]
\caption{Repository statistics across 20 repositories spanning 10 domains. \textit{\#Test fn.} and \textit{\#Test cases} indicate the number of component-level functions covered by human-curated unit tests and the total number of test cases collected from DevEval~\citep{deveval}, respectively.}
\label{tab:dataset}
\centering
\resizebox{\linewidth}{!}{%
\begin{tabular}{c l l rrrr r rr}
\toprule
\# & \textbf{Domain} & \textbf{Repository} & \textbf{Files} & \textbf{Classes} & \textbf{Methods} & \textbf{Functions} & \textbf{Components} & \textbf{\#Test fn.} & \textbf{\#Test cases} \\
\midrule
1  & \multirow{2}{*}{Communications}            & IMAPClient         &  28 &  13 &  36 &  35 &  84  &  33 &  71 \\
2  &                                            & Wikipedia-API      &   4 &   5 &  47 &   7 &  59  &  13 &  46 \\
\addlinespace[2pt]
3  & \multirow{2}{*}{Database}                  & bplustree          &  10 &   1 &  30 &   2 &  33  &  19 &  27 \\
4  &                                            & datasette          &  42 &  53 & 174 &  98 & 325  &  42 &  59 \\
\addlinespace[2pt]
5  & \multirow{2}{*}{Internet}                  & Jinja2             &  37 &  22 &  84 &  63 & 169  &  17 &  40 \\
6  &                                            & sumy               &  44 &  34 & 209 &  67 & 310  &  30 &  65 \\
\addlinespace[2pt]
7  & \multirow{2}{*}{Multimedia}                & hypertools         &  92 &  11 &  45 &  52 & 108  &  11 &  16 \\
8  &                                            & mingus             &  43 &  56 & 373 & 240 & 669  &  52 &  60 \\
\addlinespace[2pt]
9  & \multirow{2}{*}{Scientific-Engineering}    & csvkit             &  25 &  28 & 107 &  31 & 166  &   4 &  12 \\
10 &                                            & folium             &  47 &  47 &  74 &  70 & 191  &   3 &   9 \\
\addlinespace[2pt]
11 & \multirow{2}{*}{Security}                  & zxcvbn-python      &   9 &   1 &   1 &  40 &  42  &  18 &  25 \\
12 &                                            & trailscraper       &  14 &   9 &  43 &  35 &  87  &  11 &  28 \\
\addlinespace[2pt]
13 & \multirow{2}{*}{Software-Development}      & PySnooper          &   8 &  10 &  37 &  17 &  64  &   3 &   3 \\
14 &                                            & ydata-profiling    & 194 & 125 & 209 & 342 & 676  &  17 &  25 \\
\addlinespace[2pt]
15 & \multirow{2}{*}{System}                    & exodus-bundler     &  10 &  15 &  35 &  26 &  76  &  16 &  23 \\
16 &                                            & flower             &  29 &  52 & 117 &  38 & 207  &   4 &  15 \\
\addlinespace[2pt]
17 & \multirow{2}{*}{Text-Processing}           & parsel             &   8 &  11 &  49 &  19 &  79  &   3 &   3 \\
18 &                                            & online-judge-tools &  20 &  13 &  17 &  58 &  88  &   3 &  13 \\
\addlinespace[2pt]
19 & \multirow{2}{*}{Utilities}                 & mackup             &   9 &   6 &  32 &  17 &  55  &   7 &  11 \\
20 &                                            & PyJWT              &  17 &  24 &  54 &  28 & 106  &   6 &  13 \\
\midrule
\multicolumn{3}{l}{\textbf{Total}} & \textbf{690} & \textbf{536} & \textbf{1{,}773} & \textbf{1{,}285} & \textbf{3{,}594} & \textbf{312} & \textbf{564} \\
\bottomrule
\end{tabular}%
}
\end{table}

\subsection{Implementation details}~\label{B.4}

\paragraph{Backbone models.} All experiments on open-weight models are served through a vLLM server running on 2$\times$ NVIDIA RTX Pro 6000 GPUs. Closed-weight models are accessed via their official APIs. We instantiate MemDocAgent with two backbones to verify that the framework is not tied to a particular model family: Qwen3-Coder-30B-A3B (temperature: 0.1, max input tokens: 32000, max output tokens: 4096) as an open-weight backbone, and GPT-5-mini (temperature: 1.0, max input tokens: 100000, max output tokens: 4096) as a closed-weight backbone. Both backbones share the same agent scaffolding, prompts, and hyperparameters; only the underlying generator differs.

\paragraph{MemDocAgent configuration.}
External knowledge retrieval is delegated to the Perplexity API, which is invoked only when local repository evidence is insufficient. We use the Perplexity API (\url{https://www.perplexity.ai/}) for external query requests. The sub-task loop is capped at $10$ steps, the verify loop at $2$ steps, and the verify acceptance threshold is set to $0.9$. All remaining hyperparameters follow the default decoding settings of each backbone.


\section{Additional experimental results}~\label{C}

\subsection{Repository-wise experimental results}~\label{C.1}

While Section~\ref{6.1} reports the overall documentation quality averaged across the 20 repositories, Tables~\ref{tab:repo-wise-comp}, \ref{tab:repo-wise-truth}, and \ref{tab:repo-wise-help} present the per-repository breakdown of completeness, truthfulness, and helpfulness, respectively, comparing MemDocAgent against the open-source baselines. MemDocAgent consistently outperforms the baselines across all three criteria, and its advantage holds uniformly across repositories regardless of domain or scale.

\begin{table}[h]
\caption{Per-repository Completeness scores across 20 repositories. Top: Qwen3-Coder-30B-A3B-Instruct backbone. Bottom: GPT-5-mini backbone. Best score per row in \textbf{bold}.}
\label{tab:repo-wise-comp}
\centering
\resizebox{0.9\linewidth}{!}{%
\begin{tabular}{l l ccccc}
\toprule
Domain & Repository & Prompting & RepoAgent & DocAgent & CodeWiki & \textbf{MemDocAgent} \\
\midrule
\multicolumn{7}{c}{\textbf{(a) Qwen3-Coder-30B-A3B-Instruct}} \\
\midrule
\multirow{2}{*}{Communications}
  & Wikipedia-API       & 0.807 & 0.809 & 0.951 & 0.557 & \textbf{0.998} \\
  & IMAPClient          & 0.827 & 0.837 & 0.835 & 0.550 & \textbf{0.978} \\
\multirow{2}{*}{Database}
  & bplustree           & 0.892 & 0.823 & 0.892 & 0.633 & \textbf{0.974} \\
  & datasette           & 0.894 & 0.847 & 0.858 & 0.624 & \textbf{0.972} \\
\multirow{2}{*}{Internet}
  & Jinja2              & 0.847 & 0.781 & 0.855 & 0.602 & \textbf{0.985} \\
  & sumy                & 0.846 & 0.856 & 0.906 & 0.600 & \textbf{0.976} \\
\multirow{2}{*}{Multimedia}
  & hypertools          & 0.844 & 0.770 & 0.869 & 0.664 & \textbf{0.987} \\
  & mingus              & 0.800 & 0.851 & 0.777 & 0.612 & \textbf{0.971} \\
\multirow{2}{*}{Scientific-Eng.}
  & csvkit              & 0.831 & 0.787 & 0.825 & 0.676 & \textbf{0.970} \\
  & folium              & 0.839 & 0.838 & 0.599 & 0.613 & \textbf{0.966} \\
\multirow{2}{*}{Security}
  & zxcvbn-python       & 0.772 & 0.750 & 0.878 & 0.675 & \textbf{0.943} \\
  & trailscraper        & 0.832 & 0.845 & 0.873 & 0.632 & \textbf{0.990} \\
\multirow{2}{*}{Software-Dev.}
  & PySnooper           & 0.797 & 0.779 & 0.870 & 0.690 & \textbf{0.987} \\
  & ydata-profiling     & 0.855 & 0.830 & 0.815 & 0.660 & \textbf{0.989} \\
\multirow{2}{*}{System}
  & exodus-bundler      & 0.843 & 0.770 & 0.867 & 0.699 & \textbf{0.979} \\
  & flower              & 0.839 & 0.843 & 0.838 & 0.621 & \textbf{0.985} \\
\multirow{2}{*}{Text-Processing}
  & parsel              & 0.868 & 0.784 & 0.907 & 0.567 & \textbf{0.989} \\
  & online-judge-tools  & 0.856 & 0.825 & 0.862 & 0.580 & \textbf{0.989} \\
\multirow{2}{*}{Utilities}
  & mackup              & 0.803 & 0.789 & 0.796 & 0.622 & \textbf{0.962} \\
  & PyJWT               & 0.870 & 0.843 & 0.829 & 0.568 & \textbf{0.983} \\
\midrule
\multicolumn{2}{l}{\textbf{Average}}
                        & 0.838 & 0.813 & 0.845 & 0.622 & \textbf{0.979} \\
\midrule
\multicolumn{7}{c}{\textbf{(b) GPT-5-mini}} \\
\midrule
\multirow{2}{*}{Communications}
  & Wikipedia-API       & 0.630 & 0.809 & 0.925 & 0.746 & \textbf{0.979} \\
  & IMAPClient          & 0.634 & 0.770 & 0.848 & 0.780 & \textbf{0.947} \\
\multirow{2}{*}{Database}
  & bplustree           & 0.645 & 0.828 & 0.859 & 0.710 & \textbf{0.958} \\
  & datasette           & 0.615 & 0.783 & 0.881 & 0.794 & \textbf{0.952} \\
\multirow{2}{*}{Internet}
  & Jinja2              & 0.572 & 0.781 & 0.866 & 0.764 & \textbf{0.938} \\
  & sumy                & 0.568 & 0.793 & 0.914 & 0.768 & \textbf{0.973} \\
\multirow{2}{*}{Multimedia}
  & hypertools          & 0.544 & 0.770 & 0.882 & 0.676 & \textbf{0.960} \\
  & mingus              & 0.568 & 0.789 & 0.893 & 0.765 & \textbf{0.952} \\
\multirow{2}{*}{Scientific-Engineering}
  & csvkit              & 0.825 & 0.787 & 0.834 & 0.797 & \textbf{0.960} \\
  & folium              & 0.784 & 0.765 & 0.684 & 0.827 & \textbf{0.954} \\
\multirow{2}{*}{Security}
  & zxcvbn-python       & 0.330 & 0.750 & 0.896 & 0.749 & \textbf{0.902} \\
  & trailscraper        & 0.622 & 0.777 & 0.900 & 0.807 & \textbf{0.966} \\
\multirow{2}{*}{Software-Development}
  & PySnooper           & 0.558 & 0.779 & 0.864 & 0.829 & \textbf{0.959} \\
  & ydata-profiling     & 0.626 & 0.760 & 0.843 & 0.729 & \textbf{0.982} \\
\multirow{2}{*}{System}
  & exodus-bundler      & 0.669 & 0.770 & 0.717 & 0.854 & \textbf{0.955} \\
  & flower              & 0.715 & 0.776 & 0.873 & 0.646 & \textbf{0.958} \\
\multirow{2}{*}{Text-Processing}
  & parsel              & 0.646 & 0.784 & 0.901 & 0.841 & \textbf{0.967} \\
  & online-judge-tools  & 0.535 & 0.724 & 0.897 & 0.769 & \textbf{0.942} \\
\multirow{2}{*}{Utilities}
  & mackup              & 0.784 & 0.789 & 0.874 & 0.877 & \textbf{0.973} \\
  & PyJWT               & 0.763 & 0.781 & 0.839 & 0.819 & \textbf{0.974} \\
\midrule
\multicolumn{2}{l}{\textbf{Average}}
                        & 0.632 & 0.778 & 0.860 & 0.777 & \textbf{0.958} \\
\bottomrule
\end{tabular}%
}
\end{table}

\begin{table}[h]
\caption{Per-repository Truthfulness scores across 20 repositories. Top: Qwen3-Coder-30B-A3B-Instruct backbone. Bottom: GPT-5-mini backbone. Best score per row in \textbf{bold}.}
\label{tab:repo-wise-truth}
\centering
\resizebox{0.9\linewidth}{!}{%
\begin{tabular}{l l ccccc}
\toprule
Domain & Repository & Prompting & RepoAgent & DocAgent & CodeWiki & \textbf{MemDocAgent} \\
\midrule
\multicolumn{7}{c}{\textbf{(a) Qwen3-Coder-30B-A3B-Instruct}} \\
\midrule
\multirow{2}{*}{Communications}
  & Wikipedia-API       & 0.680 & 0.784 & 0.782 & 0.823 & \textbf{0.919} \\
  & IMAPClient          & 0.657 & 0.812 & 0.804 & 0.882 & \textbf{0.920} \\
\multirow{2}{*}{Database}
  & bplustree           & 0.679 & 0.838 & 0.766 & \textbf{0.926} & 0.913 \\
  & datasette           & 0.623 & 0.808 & 0.866 & 0.916 & \textbf{0.940} \\
\multirow{2}{*}{Internet}
  & Jinja2              & 0.645 & 0.811 & 0.722 & 0.542 & \textbf{0.963} \\
  & sumy                & 0.593 & 0.831 & 0.774 & 0.423 & \textbf{0.876} \\
\multirow{2}{*}{Multimedia}
  & hypertools          & 0.661 & 0.863 & 0.849 & 0.783 & \textbf{0.945} \\
  & mingus              & 0.547 & 0.801 & 0.792 & 0.781 & \textbf{0.845} \\
\multirow{2}{*}{Scientific-Eng.}
  & csvkit              & 0.657 & 0.864 & 0.880 & \textbf{0.970} & 0.950 \\
  & folium              & 0.668 & 0.847 & 0.822 & 0.831 & \textbf{0.875} \\
\multirow{2}{*}{Security}
  & zxcvbn-python       & 0.606 & 0.822 & 0.734 & 0.509 & \textbf{0.934} \\
  & trailscraper        & 0.542 & 0.864 & 0.878 & 0.310 & \textbf{0.897} \\
\multirow{2}{*}{Software-Dev.}
  & PySnooper           & 0.766 & 0.883 & 0.724 & 0.850 & \textbf{0.894} \\
  & ydata-profiling     & 0.584 & 0.837 & 0.824 & 0.852 & \textbf{0.932} \\
\multirow{2}{*}{System}
  & exodus-bundler      & 0.679 & 0.747 & 0.710 & \textbf{0.970} & 0.816 \\
  & flower              & 0.698 & 0.900 & 0.880 & 0.634 & \textbf{0.941} \\
\multirow{2}{*}{Text-Processing}
  & parsel              & 0.755 & 0.892 & 0.802 & 0.947 & \textbf{0.960} \\
  & online-judge-tools  & 0.648 & 0.814 & 0.857 & 0.782 & \textbf{0.962} \\
\multirow{2}{*}{Utilities}
  & mackup              & 0.648 & 0.841 & 0.779 & 0.872 & \textbf{0.944} \\
  & PyJWT               & 0.640 & 0.837 & 0.758 & 0.834 & \textbf{0.897} \\
\midrule
\multicolumn{2}{l}{\textbf{Average}}
                        & 0.649 & 0.835 & 0.800 & 0.772 & \textbf{0.916} \\
\midrule
\multicolumn{7}{c}{\textbf{(b) GPT-5-mini}} \\
\midrule
\multirow{2}{*}{Communications}
  & Wikipedia-API       & 0.721 & 0.858 & 0.878 & 0.752 & \textbf{0.956} \\
  & IMAPClient          & 0.787 & 0.791 & 0.847 & 0.312 & \textbf{0.943} \\
\multirow{2}{*}{Database}
  & bplustree           & 0.772 & 0.823 & 0.873 & 0.544 & \textbf{0.972} \\
  & datasette           & 0.794 & 0.881 & 0.875 & 0.578 & \textbf{0.992} \\
\multirow{2}{*}{Internet}
  & Jinja2              & 0.777 & 0.877 & 0.849 & 0.462 & \textbf{0.944} \\
  & sumy                & 0.731 & 0.815 & 0.788 & 0.623 & \textbf{0.977} \\
\multirow{2}{*}{Multimedia}
  & hypertools          & 0.671 & 0.867 & 0.826 & 0.882 & \textbf{0.929} \\
  & mingus              & 0.671 & 0.798 & 0.795 & 0.500 & \textbf{0.922} \\
\multirow{2}{*}{Scientific-Eng.}
  & csvkit              & 0.737 & 0.901 & 0.827 & 0.695 & \textbf{0.949} \\
  & folium              & 0.818 & 0.863 & 0.837 & 0.833 & \textbf{0.989} \\
\multirow{2}{*}{Security}
  & zxcvbn-python       & 0.810 & 0.864 & 0.811 & 0.537 & \textbf{0.968} \\
  & trailscraper        & 0.708 & 0.888 & 0.817 & 0.855 & \textbf{0.937} \\
\multirow{2}{*}{Software-Dev.}
  & PySnooper           & 0.764 & 0.875 & 0.838 & 0.823 & \textbf{0.958} \\
  & ydata-profiling     & 0.734 & 0.840 & 0.833 & 0.500 & \textbf{0.984} \\
\multirow{2}{*}{System}
  & exodus-bundler      & 0.718 & 0.835 & 0.783 & 0.824 & \textbf{0.964} \\
  & flower              & 0.702 & 0.825 & 0.869 & 0.839 & \textbf{0.984} \\
\multirow{2}{*}{Text-Processing}
  & parsel              & 0.750 & 0.870 & 0.896 & 0.713 & \textbf{0.942} \\
  & online-judge-tools  & 0.711 & 0.877 & 0.844 & 0.500 & \textbf{0.901} \\
\multirow{2}{*}{Utilities}
  & mackup              & 0.743 & 0.835 & 0.785 & 0.684 & \textbf{0.933} \\
  & PyJWT               & 0.707 & 0.821 & 0.754 & 0.600 & \textbf{0.895} \\
\midrule
\multicolumn{2}{l}{\textbf{Average}}
                        & 0.741 & 0.850 & 0.831 & 0.653 & \textbf{0.952} \\
\bottomrule
\end{tabular}%
}
\end{table}

\begin{table}[h]
\caption{Per-repository Helpfulness scores across 20 repositories. Top: Qwen3-Coder-30B-A3B-Instruct backbone. Bottom: GPT-5-mini backbone. Best score per row in \textbf{bold}.}
\label{tab:repo-wise-help}
\centering
\resizebox{0.9\linewidth}{!}{%
\begin{tabular}{l l ccccc}
\toprule
Domain & Repository & Prompting & RepoAgent & DocAgent & CodeWiki & \textbf{MemDocAgent} \\
\midrule
\multicolumn{7}{c}{\textbf{(a) Qwen3-Coder-30B-A3B-Instruct}} \\
\midrule
\multirow{2}{*}{Communications}
  & Wikipedia-API       & 0.474 & 0.356 & 0.613 & 0.250 & \textbf{0.686} \\
  & IMAPClient          & 0.406 & 0.390 & 0.613 & 0.212 & \textbf{0.632} \\
\multirow{2}{*}{Database}
  & bplustree           & 0.370 & 0.369 & 0.623 & 0.750 & \textbf{0.787} \\
  & datasette           & 0.481 & 0.393 & 0.609 & 0.550 & \textbf{0.682} \\
\multirow{2}{*}{Internet}
  & Jinja2              & 0.394 & 0.363 & 0.542 & 0.533 & \textbf{0.601} \\
  & sumy                & 0.568 & 0.385 & 0.529 & 0.523 & \textbf{0.692} \\
\multirow{2}{*}{Multimedia}
  & hypertools          & 0.554 & 0.362 & 0.606 & 0.583 & \textbf{0.674} \\
  & mingus              & 0.550 & 0.354 & 0.348 & 0.521 & \textbf{0.623} \\
\multirow{2}{*}{Scientific-Eng.}
  & csvkit              & 0.539 & 0.404 & 0.620 & 0.667 & \textbf{0.669} \\
  & folium              & 0.441 & 0.460 & 0.651 & 0.682 & \textbf{0.769} \\
\multirow{2}{*}{Security}
  & zxcvbn-python       & 0.397 & 0.354 & 0.610 & \textbf{0.733} & 0.723 \\
  & trailscraper        & 0.594 & 0.380 & 0.516 & 0.724 & \textbf{0.745} \\
\multirow{2}{*}{Software-Dev.}
  & PySnooper           & 0.453 & 0.366 & 0.541 & \textbf{0.667} & 0.648 \\
  & ydata-profiling     & 0.459 & 0.384 & 0.506 & 0.523 & \textbf{0.782} \\
\multirow{2}{*}{System}
  & exodus-bundler      & 0.398 & 0.435 & 0.562 & 0.500 & \textbf{0.606} \\
  & flower              & 0.301 & 0.436 & 0.516 & 0.423 & \textbf{0.612} \\
\multirow{2}{*}{Text-Processing}
  & parsel              & 0.435 & 0.393 & 0.548 & 0.667 & \textbf{0.720} \\
  & online-judge-tools  & 0.502 & 0.396 & 0.521 & 0.745 & \textbf{0.768} \\
\multirow{2}{*}{Utilities}
  & mackup              & 0.308 & 0.383 & 0.501 & 0.583 & \textbf{0.609} \\
  & PyJWT               & 0.538 & 0.454 & 0.553 & 0.521 & \textbf{0.780} \\
\midrule
\multicolumn{2}{l}{\textbf{Average}}
                        & 0.458 & 0.391 & 0.556 & 0.568 & \textbf{0.690} \\
\midrule
\multicolumn{7}{c}{\textbf{(b) GPT-5-mini}} \\
\midrule
\multirow{2}{*}{Communications}
  & Wikipedia-API       & 0.583 & 0.413 & 0.639 & 0.817 & \textbf{0.851} \\
  & IMAPClient          & 0.634 & 0.427 & 0.722 & 0.583 & \textbf{0.762} \\
\multirow{2}{*}{Database}
  & bplustree           & 0.592 & 0.402 & 0.568 & 0.650 & \textbf{0.793} \\
  & datasette           & 0.618 & 0.455 & 0.713 & 0.750 & \textbf{0.757} \\
\multirow{2}{*}{Internet}
  & Jinja2              & 0.627 & 0.421 & 0.701 & 0.650 & \textbf{0.764} \\
  & sumy                & 0.592 & 0.446 & 0.687 & 0.743 & \textbf{0.793} \\
\multirow{2}{*}{Multimedia}
  & hypertools          & 0.598 & 0.426 & 0.688 & 0.750 & \textbf{0.864} \\
  & mingus              & 0.586 & 0.406 & 0.559 & 0.667 & \textbf{0.722} \\
\multirow{2}{*}{Scientific-Eng.}
  & csvkit              & 0.618 & 0.471 & 0.637 & 0.650 & \textbf{0.733} \\
  & folium              & 0.679 & 0.505 & 0.710 & 0.750 & \textbf{0.851} \\
\multirow{2}{*}{Security}
  & zxcvbn-python       & 0.613 & 0.431 & 0.727 & 0.667 & \textbf{0.787} \\
  & trailscraper        & 0.608 & 0.434 & 0.633 & 0.833 & \textbf{0.850} \\
\multirow{2}{*}{Software-Dev.}
  & PySnooper           & 0.632 & 0.449 & 0.703 & 0.750 & \textbf{0.762} \\
  & ydata-profiling     & 0.618 & 0.460 & 0.756 & 0.733 & \textbf{0.862} \\
\multirow{2}{*}{System}
  & exodus-bundler      & 0.611 & 0.488 & 0.616 & 0.633 & \textbf{0.866} \\
  & flower              & 0.625 & 0.504 & 0.655 & 0.693 & \textbf{0.749} \\
\multirow{2}{*}{Text-Processing}
  & parsel              & 0.631 & 0.449 & 0.657 & 0.633 & \textbf{0.762} \\
  & online-judge-tools  & 0.647 & 0.463 & 0.688 & 0.717 & \textbf{0.793} \\
\multirow{2}{*}{Utilities}
  & mackup              & 0.594 & 0.460 & 0.680 & 0.750 & \textbf{0.823} \\
  & PyJWT               & 0.638 & 0.468 & 0.674 & 0.733 & \textbf{0.861} \\
\midrule
\multicolumn{2}{l}{\textbf{Average}}
                        & 0.617 & 0.449 & 0.671 & 0.708 & \textbf{0.800} \\
\bottomrule
\end{tabular}%
}
\end{table}

\subsection{Scalability analysis}~\label{C.2}

Table~\ref{tab:scalability} reports how each system's performance shifts as repository size increases. We sort the 20 repositories by file count and split them into a smaller half (Stage 1) and a larger half (Stage 2), then compare performance across the two splits.

The open-source baselines (Prompting, RepoAgent, DocAgent) consistently degrade across all four metrics, with vanilla Prompting showing the largest drop in Effectiveness (-0.086), indicating that single-shot long-context generation and stateless multi-agent decomposition both fail to capture cross-file dependencies as repositories grow. In contrast, MemDocAgent records the smallest absolute change $|\Delta|$ on every metric and even shows slight improvements on the three documentation-quality measures (Completeness +0.001, Truthfulness +0.010, Helpfulness +0.006). 
MemDocAgent maintains documentation quality even as the repository grows larger, showing that memory-guided hierarchical documentation scales reliably to large repositories.

\begin{table}[h]
\caption{Scalability analysis under Qwen3-Coder-30B-A3B-Instruct backbone: performance change from Stage 1 (10 small-sized repositories) to Stage 2 (10 large-sized repositories), with $\Delta = S_2 - S_1$. For each metric, the smallest absolute change $|\Delta|$ is in \textbf{bold}.}
\label{tab:scalability}
\centering
\resizebox{\linewidth}{!}{%
\begin{tabular}{l ccc ccc ccc ccc}
\toprule
\multirow{2}{*}{\textbf{System}}
& \multicolumn{3}{c}{Completeness}
& \multicolumn{3}{c}{Truthfulness}
& \multicolumn{3}{c}{Helpfulness}
& \multicolumn{3}{c}{Effectiveness (CodeBLEU)} \\
\cmidrule(lr){2-4} \cmidrule(lr){5-7} \cmidrule(lr){8-10} \cmidrule(lr){11-13}
& $S_1$ & $S_2$ & $\Delta$ & $S_1$ & $S_2$ & $\Delta$ & $S_1$ & $S_2$ & $\Delta$ & $S_1$ & $S_2$ & $\Delta$ \\
\midrule
Prompting & 0.620 & 0.603 & $-0.017$ & 0.761 & 0.734 & $-0.027$ & 0.625 & 0.610 & $-0.015$ & 0.423 & 0.337 & $-0.086$ \\
RepoAgent~\cite{repoagent} & 0.785 & 0.775 & $-0.010$ & 0.872 & 0.857 & $-0.015$ & 0.462 & 0.441 &  $-0.021$ & 0.313 & 0.298 & $-0.015$ \\
DocAgent~\cite{docagent}  & 0.862 & 0.857 & $-0.005$ & 0.827 & 0.814 & $-0.013$ & 0.687 & 0.645 &  $-0.042$ & 0.209 & 0.156 & $-0.053$ \\
\rowcolor[HTML]{EFEFEF}
\textbf{MemDocAgent}
  & \textbf{0.954} & \textbf{0.955} & $\mathbf{+0.001}$
  & \textbf{0.741} & \textbf{0.750} & $\mathbf{+0.010}$
  & \textbf{0.751} & \textbf{0.756} & $\mathbf{+0.006}$
  & \textbf{0.282} & \textbf{0.268} & $\mathbf{-0.014}$ \\
\bottomrule
\end{tabular}%
}
\end{table}

\subsection{Cross-language generalization on large-scale repositories}~\label{C.3}

Our framework can generate comprehensive documentation for repositories written in languages beyond Python. MemDocAgent's dependency parser is built on tree-sitter~\footnote{\url{https://tree-sitter.github.io/tree-sitter/}}, which natively supports JavaScript, TypeScript, Java, C++, C, C\#, PHP, and Kotlin in addition to Python. To further demonstrate both the language-agnosticism and the size stability analyzed in Appendix~\ref{C.2}, we collect two additional large-scale GitHub repositories: a Java repository (\texttt{elastic/logstash}; 395 files, 4{,}134 components) and a C++ repository (\texttt{electron/electron}; 835 files, 9{,}190 components). 
These repositories are 2--4$\times$ larger in file count and 6--14$\times$ larger in component count than the largest repository used in our main experiments. Table~\ref{tab:language} reports the cross-language results on these two large-scale repositories using the Qwen3-Coder backbone. In such large-scale settings, per-document time increases slightly because of the need to read dependency-related documents at the module and repository levels; nevertheless, MemDocAgent remains more efficient than the SOTA baseline, DocAgent, in both quality and time cost.

\begin{table}[h]
\caption{Cross-language results on large-scale repositories. MemDocAgent versus the prior SOTA, DocAgent, on Java (elastic/logstash) and C++ (electron/electron) using the Qwen3-Coder backbone. Best in \textbf{bold}.}
\label{tab:language}
\centering
\resizebox{\linewidth}{!}{%
\begin{tabular}{lll cccc}
\toprule
Language & Repository & Method & Completeness & Truthfulness & Helpfulness & Time/Doc(s) \\
\midrule
Java &    elastic/logstash & DocAgent &	0.835 &	0.730 &	0.600 &	72.80 \\
    & & \textbf{MemDocAgent} &	\textbf{0.978} &	\textbf{0.957} &	\textbf{0.703} &	\textbf{60.21} \\
\midrule
C++	& electron/electron & DocAgent	& 0.820	& 0.810	& 0.550	 & 69.10 \\
& & \textbf{MemDocAgent} &	\textbf{0.988} & 	\textbf{0.930} &	\textbf{0.695} &	\textbf{58.39} \\
\bottomrule

\end{tabular}%
}
\end{table}

\subsection{Details of information sufficiency}\label{C.4}

Figure~\ref{fig:sufficiency_exp} compares the average results across two code-generation models, GPT-4o-mini and Claude Haiku 4.5, and Table~\ref{tab:full_sufficiency} provides the full per-model results.

\begin{table}[h]
\caption{Information sufficiency evaluation via downstream code reproducibility}
\label{tab:full_sufficiency}
\centering
\resizebox{\linewidth}{!}{%
\begin{tabular}{ll ccc ccc ccc}
\toprule
& & \multicolumn{3}{c}{\textbf{GPT-4o-mini}} & \multicolumn{3}{c}{\textbf{Claude Haiku 4.5}} & \multicolumn{3}{c}{\textbf{Average}} \\
\cmidrule(lr){3-5} \cmidrule(lr){6-8} \cmidrule(lr){9-11}
\multicolumn{2}{l}{\textbf{Method}} & Pass@1 & Pass@3 & CodeBLEU & Pass@1 & Pass@3 & CodeBLEU & Pass@1 & Pass@3 & CodeBLEU \\
\midrule
\multicolumn{11}{c}{\textit{Reference baselines}} \\
\cmidrule(lr){1-11}
\multicolumn{2}{l}{No docs}      & 0.107 & 0.129 & 0.142 & 0.154 & 0.198 & 0.218 & 0.131 & 0.163 & 0.180 \\
\multicolumn{2}{l}{Ground-truth} & 0.151 & 0.191 & 0.221 & 0.199 & 0.253 & 0.281 & 0.175 & 0.222 & 0.251 \\
\midrule
\multicolumn{11}{c}{\textit{Backbone: Qwen3-Coder-30B-A3B-Instruct}} \\
\cmidrule(lr){1-11}
Prompting                     & C & 0.156 & 0.196 & 0.314 & 0.262 & 0.325 & 0.341 & 0.209 & 0.261 & 0.327 \\
RepoAgent                     & C & 0.231 & 0.253 & 0.290 & 0.352 & 0.357 & 0.335 & 0.292 & 0.305 & 0.312 \\
DocAgent                      & C & 0.158 & 0.168 & 0.249 & 0.305 & 0.315 & 0.289 & 0.231 & 0.242 & 0.269 \\
\rowcolor[HTML]{EFEFEF}
\cellcolor[HTML]{EFEFEF}                           & \cellcolor[HTML]{EFEFEF}C     & 0.273 & 0.297 & 0.335 & 0.535 & 0.550 & 0.403 & 0.404 & 0.424 & 0.369 \\
\rowcolor[HTML]{EFEFEF}
\cellcolor[HTML]{EFEFEF}\textbf{MemDocAgent}       & \cellcolor[HTML]{EFEFEF}C+M   & \textbf{0.354} & \textbf{0.362} & 0.376 & 0.568 & 0.575 & 0.441 & \textbf{0.461} & \textbf{0.469} & 0.409 \\
\rowcolor[HTML]{EFEFEF}
\cellcolor[HTML]{EFEFEF}                           & \cellcolor[HTML]{EFEFEF}C+M+R & 0.349 & 0.356 & \textbf{0.392} & \textbf{0.573} & \textbf{0.578} & \textbf{0.446} & \textbf{0.461} & 0.467 & \textbf{0.419} \\
\midrule
\multicolumn{11}{c}{\textit{Backbone: GPT-5-mini}} \\
\cmidrule(lr){1-11}
Prompting                     & C & 0.226 & 0.233 & 0.350 & 0.442 & 0.461 & 0.408 & 0.334 & 0.347 & 0.379 \\
RepoAgent                     & C & 0.345 & 0.359 & 0.345 & 0.352 & 0.359 & 0.351 & 0.349 & 0.359 & 0.348 \\
DocAgent                      & C & 0.265 & 0.279 & 0.175 & 0.259 & 0.263 & 0.204 & 0.262 & 0.271 & 0.190 \\
\rowcolor[HTML]{EFEFEF}
\cellcolor[HTML]{EFEFEF}                           & \cellcolor[HTML]{EFEFEF}C     & 0.390 & 0.395 & 0.373 & 0.448 & 0.449 & 0.472 & 0.419 & 0.422 & 0.423 \\
\rowcolor[HTML]{EFEFEF}
\cellcolor[HTML]{EFEFEF}\textbf{MemDocAgent}       & \cellcolor[HTML]{EFEFEF}C+M   & 0.401 & 0.411 & 0.376 & 0.480 & 0.505 & \textbf{0.479} & 0.441 & 0.458 & \textbf{0.428} \\
\rowcolor[HTML]{EFEFEF}
\cellcolor[HTML]{EFEFEF}                           & \cellcolor[HTML]{EFEFEF}C+M+R & \textbf{0.428} & \textbf{0.430} & \textbf{0.379} & \textbf{0.509} & \textbf{0.553} & 0.477 & \textbf{0.469} & \textbf{0.492} & \textbf{0.428} \\
\bottomrule
\end{tabular}%
}
\end{table}

\subsection{Memory architecture ablation}~\label{C.10}
Unlike generic text or embedding memory, each RepoMemory record stores decomposed atomic claims, verification scores, dependency relations, and child units. This schema enables claim-level cross-document verification at commit time. Moreover, by admitting only VERIFY-passed documents as trusted references, RepoMemory acts as a consistency-enforcement mechanism rather than a passive store, preventing low-quality documents from propagating errors.

To isolate the effect of this design, we compare against a baseline that replaces RepoMemory with a generic vector memory (ReAct + embedding top-1 retrieval, without schema or verification gating). As shown in Table~\ref{tab:memory_ablation}, adding generic vector memory yields only marginal or even negative changes, since it injects unverified, inconsistent context with no consistency control. In contrast, the full RepoMemory design produces substantial gains across all metrics, confirming that the improvement stems from our documentation-specific memory design, not from the mere presence of memory.

\begin{table}[h]
\caption{Ablation study of the memory mechanism. We compare our full method against variants without memory and with vector-based memory. Best in \textbf{bold}.}
\label{tab:memory_ablation}
\centering
\resizebox{\linewidth}{!}{%
\begin{tabular}{l ccc ccc}
\toprule
& \multicolumn{3}{c}{Documentation Quality} & \multicolumn{3}{c}{Information Sufficiency} \\
\cmidrule(lr){2-4} \cmidrule(lr){5-7}
Method & Completeness & Truthfulness & Helpfulness & Pass@1 & Pass@3 & CodeBLEU \\
\midrule
Ours w/o memory        & 0.903 & 0.768 & 0.593 & 0.324 & 0.333 & 0.190 \\
Ours w/ vector memory  & 0.910 & 0.760 & 0.480 & 0.332 & 0.350 & 0.280 \\
\textbf{Ours (MemDocAgent)}          & \textbf{0.979} & \textbf{0.916} & \textbf{0.690} & \textbf{0.461} & \textbf{0.467} & \textbf{0.419} \\
\bottomrule
\end{tabular}%
}
\end{table}

\subsection{Traversal order ablation}~\label{C.6}
To verify the effectiveness of our traversal order design, we run an ablation that randomizes the traversal order (Shuffled) and compare it against our dependency- and hierarchy-aware ordering (Ordered).
The key point is that the traversal order itself forms a cumulative context: each unit is documented on top of the documentation already produced for its dependencies and child units. If each unit were an independent short-horizon task, shuffling the order would leave the results unchanged. As shown in Table~\ref{tab:ordering_ablation}, however, disrupting the order degrades every metric. The drop is largest on Truthfulness (-0.198), the dimension governed by cross-document consistency, since shuffling removes the verified dependency documents needed at verification time. This confirms that the traversal order acts as a cumulative context rather than a mere processing sequence, providing direct evidence that MemDocAgent performs cumulative, dependency-grounded execution rather than merely processing a long sequence of independent tasks.

\begin{table}[h]
\caption{Effect of traversal ordering. We compare our dependency- and hierarchy-aware ordering (Ordered) against a randomized order (Shuffled) on 10 repositories with the Qwen3-Coder backbone, where $\Delta = \text{Shuffled} - \text{Ordered}$. Best in \textbf{bold}.}
\label{tab:ordering_ablation}
\centering
\resizebox{\linewidth}{!}{%
\begin{tabular}{ccc ccc ccc ccc}
\toprule
\multicolumn{3}{c}{Completeness}
& \multicolumn{3}{c}{Truthfulness}
& \multicolumn{3}{c}{Helpfulness}
& \multicolumn{3}{c}{Effectiveness (CodeBLEU)} \\
\cmidrule(lr){1-3} \cmidrule(lr){4-6} \cmidrule(lr){7-9} \cmidrule(lr){10-12}
Ordered & Shuffled & $\Delta$
& Ordered & Shuffled & $\Delta$
& Ordered & Shuffled & $\Delta$
& Ordered & Shuffled & $\Delta$ \\
\midrule
\textbf{0.954} & 0.930 & $-0.024$
& \textbf{0.741} & 0.543 & $-0.198$
& \textbf{0.751} & 0.635 & $-0.016$
& \textbf{0.282} & 0.250 & $-0.032$ \\
\bottomrule
\end{tabular}%
}
\end{table}

\subsection{Prompt controlled comparisons}~\label{C.7}
A remaining concern is that MemDocAgent's gains might stem from its output format rather than its core design. To rule this out, we re-evaluate all baselines at the component level using the same output template as MemDocAgent (templated), fully aligning the output format across methods. As shown in Table~\ref{tab:templated}, applying the shared template raises every baseline slightly, yet MemDocAgent still leads on all three metrics by a clear margin. This confirms that the improvement comes from the proposed memory and verification mechanisms, not from the output format.

\begin{table}[h]
\caption{Comparison of original and templated documentation across metrics. For each metric, we report the original and templated scores. Best in \textbf{bold}.}
\label{tab:templated}
\centering
\resizebox{\linewidth}{!}{%
\begin{tabular}{l cc cc cc}
\toprule
\multirow{2}{*}{Method}
& \multicolumn{2}{c}{Completeness}
& \multicolumn{2}{c}{Truthfulness}
& \multicolumn{2}{c}{Helpfulness} \\
\cmidrule(lr){2-3} \cmidrule(lr){4-5} \cmidrule(lr){6-7}
& original & templated & original & templated & original & templated \\
\midrule
Prompting            & 0.886 & 0.904 & 0.694 & 0.701 & 0.577 & 0.581 \\
RepoAgent~\cite{repoagent} & 0.813 & 0.890 & 0.835 & 0.850 & 0.391 & 0.400 \\
DocAgent~\cite{docagent}   & 0.845 & 0.880 & 0.800 & 0.857 & 0.556 & 0.576 \\
\textbf{MemDocAgent}  & \textbf{0.987} & \textbf{0.987} & \textbf{0.916} & \textbf{0.916} & \textbf{0.691} & \textbf{0.916} \\
\bottomrule
\end{tabular}%
}
\end{table}

\subsection{Hyperparameter Sensitivity}~\label{C.8}
Our system uses three fixed hyperparameters, max 2 revisions per unit, max 10 steps per sub-trajectory, and verify-pass score 0.9. We additionally conduct a sensitivity analysis measuring the trade-off among documentation quality, time cost, and termination behavior as each is varied. Table~\ref{tab:hyperparam} shows the result.

\begin{table}[h]
\caption{Hyperparameter sensitivity analysis on the Qwen3-Coder backbone. We vary (a) the maximum number of revisions, (b) the maximum steps per sub-trajectory, and (c) the \textsc{Verify}-pass threshold. Default settings are marked, and best results per metric are in \textbf{bold}.}
\label{tab:hyperparam}
\centering

\begin{subtable}{\linewidth}
\centering
\caption{Effect of the maximum number of revisions.}
\label{tab:hyper_revisions}
\resizebox{\linewidth}{!}{%
\begin{tabular}{c ccc c c}
\toprule
Max revisions & Completeness & Truthfulness & Helpfulness & Time/Doc (s) & Below-threshold termination \\
\midrule
1            & 0.842 & 0.892 & 0.684 & \textbf{50.27} & 11.8\% \\
2 (default)  & 0.979 & 0.916 & 0.690 & 53.16 & 0.08\% \\
3            & \textbf{0.980} & \textbf{0.930} & \textbf{0.695} & 61.67 & \textbf{0.02\%} \\
\bottomrule
\end{tabular}%
}
\end{subtable}

\vspace{1em}

\begin{subtable}{\linewidth}
\centering
\caption{Effect of the maximum number of steps per sub-trajectory.}
\label{tab:hyper_steps}
\resizebox{\linewidth}{!}{%
\begin{tabular}{c ccc c c}
\toprule
Max steps & Completeness & Truthfulness & Helpfulness & Time/Doc (s) & Unfinished rate \\
\midrule
5             & 0.840 & 0.608 & 0.470 & \textbf{40.21} & 20.72\% \\
10 (default)  & \textbf{0.979} & \textbf{0.916} & \textbf{0.690} & 53.16 & 0.08\% \\
15            & 0.979 & 0.916 & 0.682 & 71.24 & \textbf{0.01\%} \\
\bottomrule
\end{tabular}%
}
\end{subtable}

\vspace{1em}

\begin{subtable}{\linewidth}
\centering
\caption{Effect of the \textsc{Verify}-pass threshold.}
\label{tab:hyper_threshold}
\resizebox{\linewidth}{!}{%
\begin{tabular}{c ccc c c}
\toprule
Threshold & Completeness & Truthfulness & Helpfulness & Time/Doc (s) & Avg. revisions \\
\midrule
0.85          & 0.970 & 0.902 & 0.683 & \textbf{46.45} & \textbf{1.2} \\
0.9 (default) & \textbf{0.979} & 0.916 & 0.690 & 53.16 & 1.47 \\
0.95          & 0.975 & \textbf{0.926} & \textbf{0.695} & 68.23 & 1.9 \\
\bottomrule
\end{tabular}%
}
\end{subtable}

\end{table}

\subsection{Efficiency analysis}~\label{C.5}
We conduct time and cost efficiency analysis through Table~\ref{tab:throughput-cost}. On both Qwen3-Coder-30B-A3B-Instruct and GPT-5-mini, MemDocAgent achieves a lower average generation time per document than DocAgent (53.16s vs.60.86s and 56.72s vs.69.89s, respectively), while producing the most complete documentation hierarchy among all baselines (3,323 and 3,497 documents covering all three levels). 
API cost per document on GPT-5-mini is also marginally lower than DocAgent (\$0.0215 vs.\$0.0219).

\begin{table}[h]
\caption{Document generation time and cost analysis across baselines. Component, Module, and Repo report the number of documents produced at each level of the documentation hierarchy. \textbf{Time/doc} the average per-document generation time and \textbf{Cost/doc} the average API cost per document.}
\label{tab:throughput-cost}
\centering
\setlength{\tabcolsep}{4pt}
\resizebox{\linewidth}{!}{%
\begin{tabular}{l ccc c cc cc}
\toprule
Baseline & Component & Module & Repo & \textbf{Total docs} & Runtime (s) & \textbf{Time/doc (s)} & Total cost (\$) & \textbf{Cost/doc (\$)} \\
\midrule
\multicolumn{9}{c}{\textit{Generation model: Qwen3-Coder-30B-A3B-Instruct}} \\
\midrule
ChatLLM              & 3,226 & 77 & 20 & 3,323 & 39,780 & 11.97 & -- & -- \\
RepoAgent            & 3,650 &  0 &  0 & 3,650 & 11,340 & 3.11 & -- & -- \\
DocAgent             & 3,198 &  0 &  0 & 3,198 & 194,640 & \textbf{60.86} & -- & -- \\
CodeWiki             &       0 & 60 & 20 & 80 & 7,740 & 96.75 & -- & -- \\
\rowcolor[HTML]{EFEFEF} 
\textbf{MemDocAgent} & 3,226 & \multicolumn{1}{c}{\cellcolor[HTML]{EFEFEF}77} & \multicolumn{1}{c}{\cellcolor[HTML]{EFEFEF}20} & \multicolumn{1}{c}{\cellcolor[HTML]{EFEFEF}3,323} & \multicolumn{1}{c}{\cellcolor[HTML]{EFEFEF}176,640} & \textbf{53.16} & \multicolumn{1}{c}{\cellcolor[HTML]{EFEFEF}--}  & -- \\
\midrule
\multicolumn{9}{c}{\textit{Generation model: GPT-5-mini}} \\
\midrule
ChatLLM              & 3,226 & 77 & 20 & 3,323 & 100,860 & 30.35 & 30.35 & 0.0091 \\
RepoAgent            & 3,650 &  0 &  0 & 3,650 & 21,840 & 5.98 & 5.98 & 0.0016 \\
DocAgent             & 3,198 &  0 &  0 & 3,198 & 223,500 & \textbf{69.89} & 69.89 & \textbf{0.0219} \\
CodeWiki             &       0 & 60 & 20 & 80 & 3,900 & 48.75 & 48.75 & 0.6094 \\
\rowcolor[HTML]{EFEFEF} 
\textbf{MemDocAgent} & 3,410 & 67 & 20 & 3,497 & 198,360 &	\textbf{56.72} & 75.25 & \textbf{0.0215} \\ \bottomrule
\end{tabular}}
\end{table}

\subsection{More examples of generated documentation}~\label{C.9}
Figures~\ref{fig:ex_case1} and~\ref{fig:ex_case2} show example screenshots of documentation generated by MemDocAgent.

\begin{figure}[t]
\centering
\includegraphics[width=\linewidth]{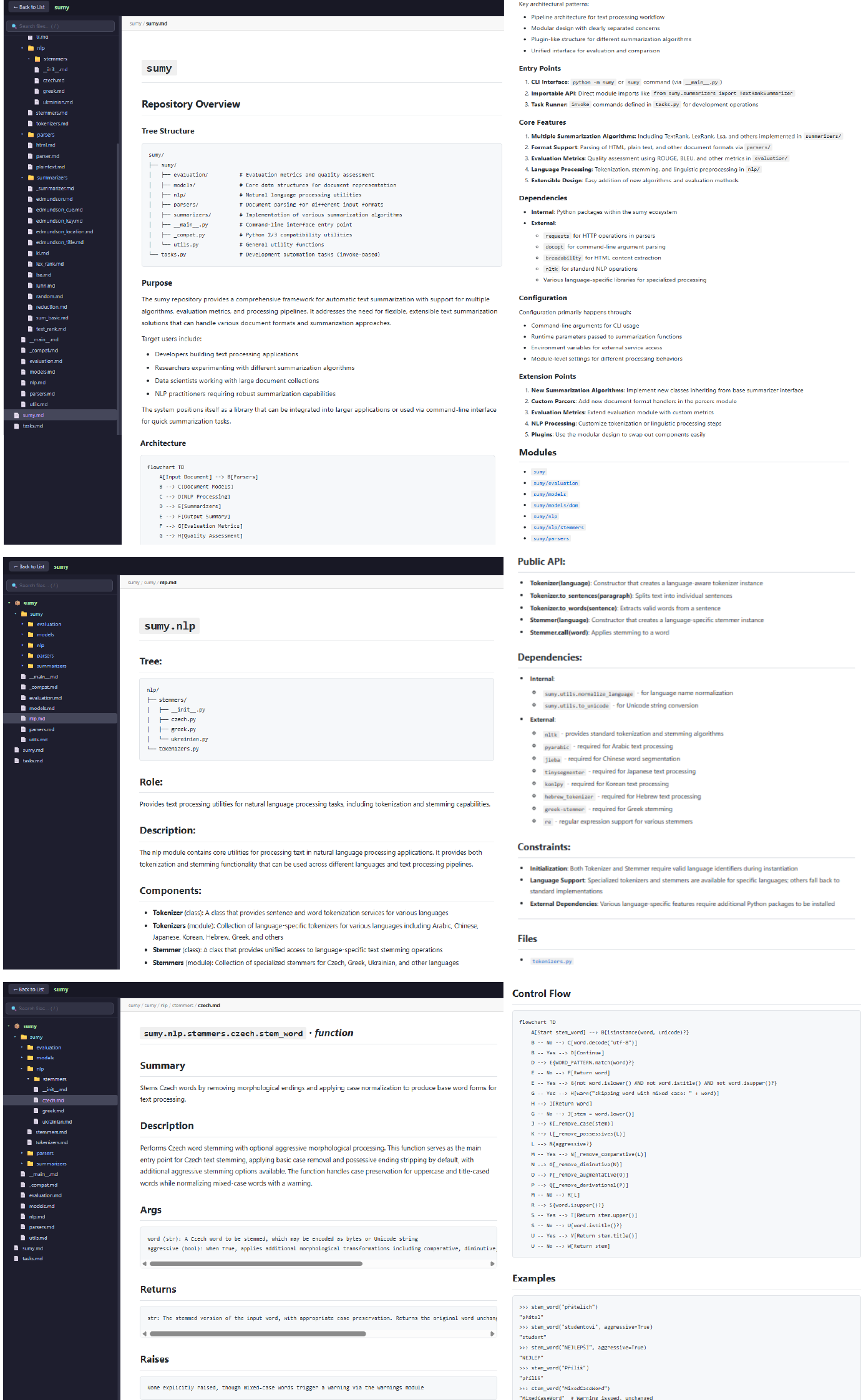}
\caption{Repo, Module, Component-level documentation example on \texttt{Sumy} Repository}
\label{fig:ex_case1}
\end{figure}

\begin{figure}[t]
\centering
\includegraphics[width=\linewidth]{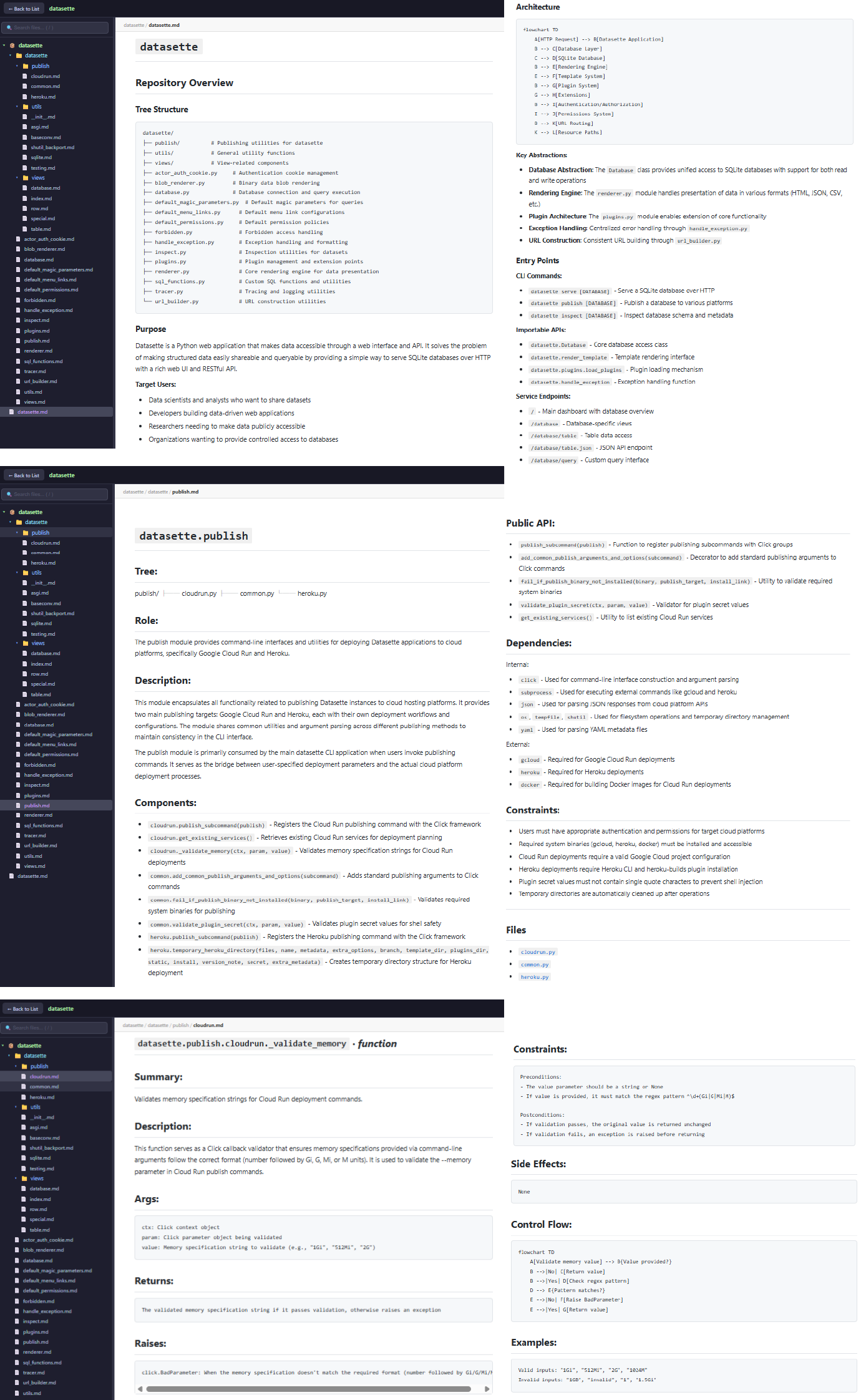}
\caption{Repo, Module, Component-level documentation example on \texttt{Datasette} Repository}
\label{fig:ex_case2}
\end{figure}


\end{document}